\documentclass[10pt,twocolumn]{article}

\usepackage[
    a4paper,
    top={26mm},
    headheight={12pt},
    headsep={5.15mm},
    text={160mm,216mm},
    marginparsep=5mm,
    marginparwidth=12mm,
    bindingoffset=6mm,
    footskip=10.13mm]{geometry}
\usepackage{graphicx}
\usepackage{authblk}
\usepackage{abstract}
\usepackage{amsmath,amssymb,amsfonts}
\usepackage{lmodern}
\usepackage[hidelinks,colorlinks=true,linkcolor=blue,citecolor=blue,urlcolor=blue]{hyperref}
\usepackage[sorting = none, backend = biber, style=numeric-comp,giveninits=true,url=false,doi=false]{biblatex}
\usepackage[all]{hypcap} 
\usepackage[labelfont=bf]{caption}
\usepackage{indentfirst,bm,siunitx,tabularx}
\usepackage{titlesec}
\usepackage[capitalise, nameinlink]{cleveref}
\usepackage[title]{appendix}
\usepackage{xcolor}
\usepackage{easyReview}

\newcommand*{\sfref}[2]{\hyperref[#1]{Supplementary Figure~\ref*{#1}#2}}
\newcommand*{\sfequ}[2]{\hyperref[#1]{#2}}

\DeclareNameAlias{default}{family-given} 
 
\renewbibmacro{in:}{}
\DeclareFieldFormat{pages}{#1}
\DeclareFieldFormat[article]{title}{#1}
\DeclareFieldFormat[inproceedings]{title}{#1}
\DeclareFieldFormat[book]{title}{#1}
\DeclareFieldFormat[misc]{title}{#1}
\DeclareFieldFormat{labelnumberwidth}{#1.}
\DeclareBibliographyDriver{article}{\usebibmacro{author}
  \newunit\addspace
  \usebibmacro{title} 
  \newunit\addspace
  \printfield{journaltitle} 
  \addspace
  \mkbibbold{\printfield{volume}} 
  \newunit\addspace
  \printfield{pages} 
  \setunit
  (\printfield{year}) 
  \finentry
}

\DeclareBibliographyDriver{inproceedings}{\usebibmacro{author} 
  \newunit\addspace
  \usebibmacro{title} 
  \newunit\addspace
  \printfield{booktitle} 
  \addspace
  \mkbibbold{\printfield{volume}} 
  \newunit\addspace
  \printfield{pages} 
  \setunit
  (\printfield{year})
  \finentry
}

\title{Neuromorphic detection and cooling of microparticles in arrays}

\author[1]{Yugang Ren\thanks{\href{mailto:yugang.ren@kcl.ac.uk}{yugang.ren@kcl.ac.uk}}}
\author[2]{Benjamin Siegel}
\author[1]{Ronghao Yin}
\author[1]{Qiongyuan Wu}
\author[1]{Jonathan Pritchett}
\author[1]{Muddassar Rashid}
\author[1,3]{James Millen\thanks{\href{mailto:james.millen@kcl.ac.uk}{james.millen@kcl.ac.uk}}}

\affil[1]{Department of Physics, King's College London, Strand, London, WC2R 2LS, United Kingdom}
\affil[2]{Wright Laboratory, Department of Physics, Yale University, New Haven,Connecticut, 06520, USA}
\affil[3]{London Centre for Nanotechnology, Department of Physics, King's College London, Strand, London, WC2R 2LS, United Kingdom}
\date{(\today)}

\begin{document}

\twocolumn[
    \maketitle
    \vspace{-1.5\baselineskip}
    \begin{onecolabstract}
    \vspace{-1\baselineskip}
        Micro-objects levitated in a vacuum are an exciting platform for precision sensing due to their low dissipation motion and the potential for control at the quantum level. Arrays of such sensors would offer increased sensitivity, directionality, and in the quantum regime the potential to exploit correlation and entanglement. We use neuromorphic detection via a single event based camera to record the motion of an array of levitated microspheres. We present a scalable method for arbitrary multiparticle tracking and control by implementing real-time feedback to {simultaneously cool the motion of three uncoupled microscale objects}.
    \end{onecolabstract}
    \vspace{\baselineskip}
]
\saythanks

\subsection*{Introduction}\label{sec1}

Modern technology relies on mechanical sensors, from accelerometers in mobile devices \cite{Shaeffer2013} to wearable health monitors \cite{Chircov2022}. As sensors are miniaturized, their surface-to-volume ratio increases and they dissipate more energy via their thermal contacts and through surface strain \cite{Imboden2014}, limiting their performance. By levitating nano- or micro-particles under ultra-high vacuum conditions, using optical, electrical or magnetic fields \cite{Millen2020rev, G-Ballestero2021rev}, one creates a mechanical oscillator with remarkably low dissipation \cite{dania2024ultrahigh, Moore2021rev}. Force sensitivities of yoctonewtons \cite{Liang2022, ranjit2016zeptonewton} and torque sensitivities at the $10^{-27}\,\textrm{N}\,\textrm{m}\,\textrm{Hz}^{-1/2}$ level \cite{ahn2020ultrasensitive} have been achieved, with levitating sensors achieving quality factors in excess of $10^{10}$ \cite{dania2024ultrahigh}, motivating researchers to use these systems to search for dark matter \cite{afek2022coherent,yin2022experiments,Moore2021rev} and gravitational waves \cite{arvanitaki2013detecting, Winstone2022}.

The control of levitated particles allows the exploration of a wide range of fundamental science \cite{Millen2020rev, G-Ballestero2021rev}, and the demonstration of cooling to the ground state of an optical potential \cite{delic2020cooling, tebbenjohanns2021quantum, magrini2021real, kamba2022optical, Piotrowski2023,ranfagni2022two} opens the door to macroscopic quantum physics \cite{millen2020quantum, barker2010, chang2010, romeroisart2010}. An emerging frontier in this field is the study of arrays of particles, which in the quantum regime would allow generation of entanglement \cite{rudolph2022force} and tests of quantum gravity \cite{bose2017spin,marletto2017gravitationally}. Interactions have been observed between pairs of levitated nanoparticles in optical \cite{arita2018optical, rieser2022tunable, livska2023cold, penny2023sympathetic}, electrodynamic (Paul) \cite{bykov20233d} and magnetic \cite{Slezak2019} traps. Detecting and controlling multiple particles in vacuum has so far involved either single particle control with sympathetic cooling \cite{arita2022all,penny2023sympathetic,bykov20233d} or small arrays of optical traps \cite{vijayan2023scalable, Hupfl2023}. Some applications will require the control of arrays of tens, or even thousands, of levitated particles \cite{afek2022coherent}.

\begin{figure}[t!] 
\centering
\includegraphics[width=0.475\textwidth]{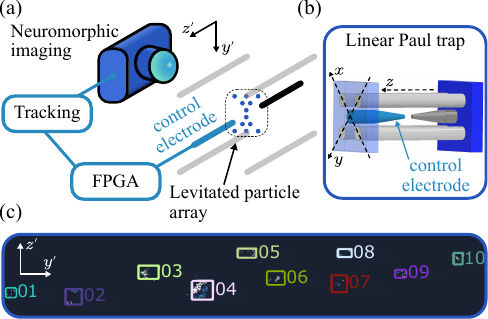}
\caption{\textbf{Neuromorphic detection of levitated particles.} (a) Schematic of the experimental setup: neuromorphic imaging via an event based camera (EBC) tracks the positions of particles levitated in an array by a linear Paul trap (four grey electrodes, {black and blue endcap electrodes}). An FPGA system processes this data to generate a feedback signal which is applied to a {control} electrode (blue) near the particle array. {(b) A schematic of the linear Paul trap, including the coordinate axes $\{x,y,z\}$ for the levitated particles, in contrast to the imaging coordinates $\{y',z'\}$ in Fig.~\ref{fig:minion}(a). (c)} An EBC image of an array of 10 particles. The EBC identifies each particle and tracks its motion (within the marked coloured box), assigning each particle an individual object id (coloured numbers) which tags the streamed data. Each object is represented by only a few pixels (white for increasing intensity, blue for decreasing intensity) making the volume of streamed data volume very low.}
\label{fig:minion}
\end{figure}

We use neuromorphic imaging for the control of arbitrary particle arrays across a wide field-of-view. Neuromorphic sensors are highly efficient detectors which mimic neurobiological information gathering \cite{Ambrogio2018, roy2019towards, Xiao2020, Liao2021, Sharma2024}. Dynamic vision sensors (DVS) are neuromorphic sensors which mimic the retinal response \cite{Vanarse2016}, detecting changes across a threshold on each pixel in an array asynchronously to produce a stream of events \cite{lichtsteiner2008128} ideally suited for object tracking \cite{Ni2015}. Together with event-driven processing algorithms \cite{nguyen2019real, rebecq2019high, cannici2019asynchronous}, DVS can achieve microsecond temporal resolution, sub-millisecond latency and high dynamic range detection (\textgreater$\,120\,\textrm{dB}$) with minimal data output at low power consumption \cite{roy2019towards, gallego2020event, Sharma2024}. Therefore, DVS are highly suited to high-speed and real-time applications requiring low-power in environments with uncontrolled light levels such as in robotics \cite{zhu2023neuromorphic}, autonomous driving \cite{gehrig2024low} and space flight \cite{Sikorski2021}, as well as finding applications in microscopy \cite{Ni2012, mangalwedhekar2023achieving} and astronomy \cite{Ng2022}. In this work, we use an event based camera (EBC) with integrated DVS to monitor the motion of an array of levitated particles with a bandwidth high enough to demonstrate real-time simultaneous feedback control of multiple {uncoupled} particles in an array.

We implement cold damping feedback \cite{Poggio2007} to cool the motion of the levitated particles \cite{Tebbenjohanns2019, Conangla2019}, a technique with demonstrated quantum ground state cooling capabilities \cite{tebbenjohanns2021quantum, magrini2021real, kamba2022optical}. We cool a single microsphere to {a temperature of a few Kelvin} and single degrees-of-freedom of multiple particles. In this work, the number of objects we can simultaneously control is only limited by the number of output channels in our feedback electronics. This single-device method for cooling and controlling particles in arrays is readily scalable due to the low data output of neuromorphic detection. Arrays of cooled micro-sensors will lead to enhanced signal-to-noise sensing through sensor fusion \cite{Palit1997,Sasiadek2002,Nazarahari2021}, enable force gradient sensing \cite{rudolph2022force}, and provide a larger interaction area without increasing the mass of the sensor \cite{afek2022coherent}. Due to the low power consumption of neuromorphic detectors our presented methods are ideal for integration into chip-scale technology \cite{melo2024vacuum}.

\subsection*{Results}
\subsubsection*{Neuromorphic imaging of levitated particle arrays}\label{sec2}

We levitate arrays of charged $5\,\mu$m diameter silica microspheres in a linear Paul trap under vacuum conditions \cite{Kane2010, Kuhlicke2014, Millen2015, Delord2017} (see Methods) and record their motion using an event based camera (EBC) \cite{ren2022event}, see Fig.~\ref{fig:minion}(a). The charged particles form a stable array due to the Coulomb repulsion between them. {Our particular Paul trap geometry (Fig.~\ref{fig:minion}(b)), and the particles' distribution of charge, means that our particle arrays are non-uniform}.

The EBC uses a neuromorphic DVS, which is an array of independent pixels featuring contrast detectors which output an event \cite{roy2019towards} in response to light levels on the pixel crossing a user-defined threshold. Pixels which do not experience the required level-change output no signal, removing the data-redundancy present in conventional cameras \cite{roy2019towards}, while allowing the full sensor to be used at all times, sometimes referred to as a dynamic region-of-interest. The EBC hardware bundles asynchronous events into equal-length frames, and uses filters to identify objects within its field-of-view \cite{Ni2015}, after which a proprietary generic tracking algorithm (GTA) tracks the motion of each object independently. We have previously demonstrated single-particle object tracking with an SNR above 35\,dB, and for a more detailed analysis of EBC performance in the context of levitated microparticles see \cite{ren2022event}. 

The EBC allows the tracking of multiple objects with a high bandwidth and a linear scaling in data output with the number of tracked objects, as compared to a rapid increase in data volume with increased region-of-interest in a conventional camera. {In our system, with fixed magnification, tracking a single particle at 1\,kHz using the entire field-of-view ($3.75\,$mm$^2$) of the EBC uses $\sim 100$\,kB\,s$^{-1}$ as compared to 64,800\,kB\,s$^{-1}$ using a standard CMOS camera (Thorlabs CS165MU/M) at the same frame rate and field-of-view}. This means that the EBC can track many hundreds of particles before the data volume becomes comparable to standard camera technology. By not having to restrict the region-of-interest, the EBC can track objects dispersed over several hundred micrometres whilst retaining high spatial resolution \cite{ren2022event}. {The low data volume leads to a correspondingly low power consumption, see Supplemental Materials S5 for more details.}

In Fig.~\ref{fig:minion}(c) we show an example of a single neuromorphic sensor being used to detect and track 10 levitated particles simultaneously.
The EBC identifies each object and tracks it in 2D (illustrated by the coloured boxes), assigning each one a stable identification number allowing us to process the position data of each particle independently. {The linear Paul trap defines the coordinate system $\{x,y,z\}$ for the levitated particles, Fig.~\ref{fig:minion}(b). The image on the EBC has a coordinate system $\{y',z'\}$, Fig.~\ref{fig:minion}(a). The $\{x,y\}$ axes are projected at $45^\circ$ onto the $y'-$axis, and the $z'-$ and $z-$axes are parallel, see Supplementary Materials S1. This projection allows us to detect all three axes of motion of the levitated particles.}

\begin{figure}[t] 
\centering
\includegraphics[width=0.47\textwidth]{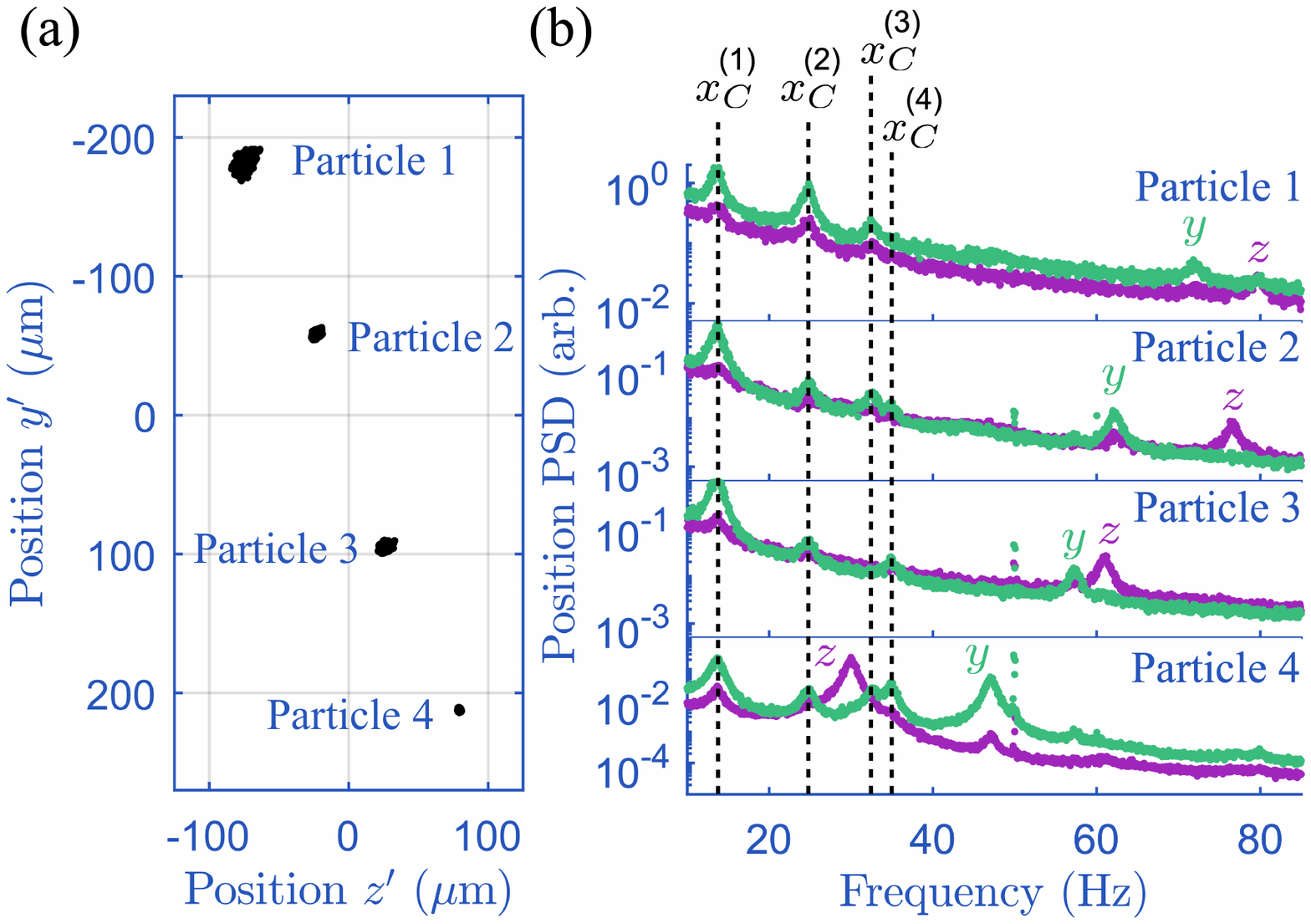}
\caption{\textbf{Real-time tracking of multiple levitated particles.} (a) Reconstructed 2D position of four levitated microparticles tracked in real-time at a 1\,kHz framerate. The amplitude of motion depends on the charge of each particle and where it sits within the levitating potential. Particles 2 and 4 are closest to and furthest from the Paul trap centre {(the centre of the coordinate axes)} respectively. (b) Position PSDs reconstructed from the 2D position tracking of the four particles in (a) (green trace: $y'-$direction, purple trace: $z'-$direction). Each particle is independently tracked. {We observe interactions between particles in the array: $x_\text{C}^{(1)}$, $x_\text{C}^{(2)}$, $x_\text{C}^{(3)}$ and $x_\text{C}^{(4)}$ are four collective modes of all the four particles in the $x$-direction. We also see the individual bare modes in the $y$- and $z$-directions. For information on identifying modes see Supplementary Materials S2 \& S3.}}
\label{fig:PSD}
\end{figure}

In Fig.~\ref{fig:PSD}(a) we reconstruct the motion and relative position of four levitated particles obtained from the 1\,kHz tracking algorithm of the EBC, which can be accessed in real-time. Particles levitated by a Paul trap undergo harmonic motion. In Fig.~\ref{fig:PSD}(b) we generate the power spectral density (PSD) from the motion of each particle in the array of four to analyze their motion in frequency space. Each particle has a different charge-to-mass ratio {(see Supplementary Materials S3)} and is levitated in a different part of the confining field, meaning each particle has different frequency modes of oscillation, which are well separated under vacuum conditions. Below, we use this fact to independently cool the motion of multiple particles. 

{The four particles are aligned along the $x-$axis. Motion of the charged particles in this direction leads to coupling between them via the Coulomb interaction \cite{Slezak2019, Svak2021, penny2023sympathetic, bykov20233d}. This is evident via the collective modes $x_C$ seen in Fig.~\ref{fig:PSD}(b). By controlling the separation between the particles, we can control the coupling strength, see Supplementary Materials S2 for further details. For the multi-particle cooling presented below, we work in a regime where the coupling between the modes is too small to measure, and perform cooling along the $z-$direction.} 

The signal-to-noise for the different particles varies due to non-uniform illumination and varying coupling to electronic noise. {It ranges from 3-30, with exact values given in Supplementary Materials S3}. All data in this work is taken at gas pressures of 2.0-$4.5\times 10^{-2}\,$mbar unless otherwise stated.

\subsubsection*{Single particle cold damping using neuromorphic imaging}\label{sec3}

\begin{figure}[htbp!]
\centering 
\includegraphics[width=0.475\textwidth]{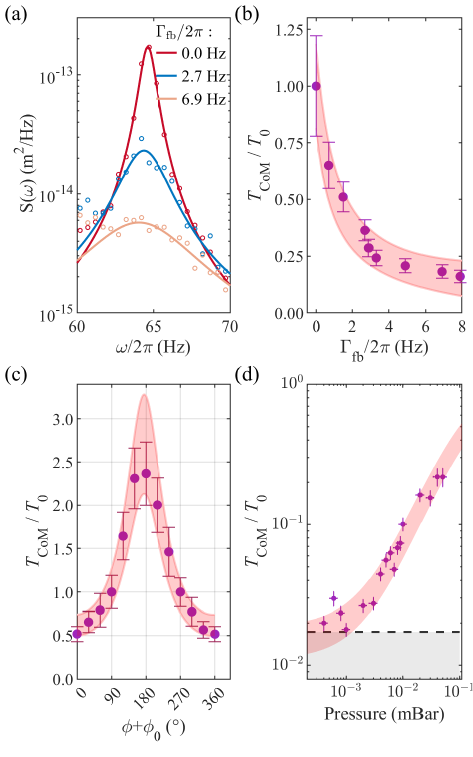}
\caption{\textbf{Single particle cooling.} (a) A selection of single particle PSDs of the {measured }motion along the $z-$axis with different feedback gains ($\mathit{\Gamma_{\rm fb}}/2\pi$), fit with equation~(\ref{eq:PSD}). (b) Extracted $T_{\text{CoM}}$ relative to initial temperature $T_0$ {over a wider range of feedback gains than presented in (a)}. (c) Effect of the feedback loop phase delay on the temperature of the particle. (d) {Variation in single particle temperature with fixed feedback gain and phase, as the background gas pressure decreases. When the pressure reaches $10^{-3}$\,mbar we cool to the noise floor of our system, as indicated by the grey shaded region, corresponding to $T_{\textrm{CoM}}=(6.8 \pm 0.7)\,\textrm{K}$. Subfigures (b-d) include the model in equation~(\ref{eq:Coolingphase}) with the parameters extracted by fitting equation~(\ref{eq:PSD}) to the PSDs of the measured particle motion - the pink shaded areas represent the uncertainty in these parameters. All experimental error-bars in the figure are derived by taking 15 repeat experiments at each set of parameters to calculate a mean and standard deviation.}}
\label{fig:single-particle cooling}
\end{figure}

There are many reasons why it is desirable to control the energy of a levitated particle. Although the sensitivity of a levitated sensor does not increase through cooling \cite{monteiro2020force}, rapid damping of the motion increases the stability and measurement bandwidth of the system. Reduction of the particle energy to the ground-state of the levitating potential \cite{delic2020cooling, magrini2021real,piotrowski2023simultaneous} opens up a toolbox of quantum control \cite{G-Ballestero2021rev} and sensitivity enhancement \cite{Mason2019, Cernotik2020} {mechanisms}. 

Cold damping is a feedback method whereby a force proportional to the velocity of an oscillator opposes its motion. Depending on the phase of the feedback force relative to the motion, this method {damps (cools) or amplifies (heats) the oscillator. When cooling, input and output noise of the feedback electronics limits the ultimate temperature.}

We process the position data from the EBC using a field programmable gate array (FPGA) to generate a feedback signal proportional to velocity, with variable gain and phase. This signal is filtered and applied to {a control electrode}, see Fig.~\ref{fig:minion}(a).

Figure~\ref{fig:single-particle cooling}(a) shows the PSD of a particle's {measured} motion along the $z-$axis as it is cooled via cold damping. {The shape of the PSD from the in-loop detector is given by \cite{Poggio2007,Conangla2019}}:{
\begin{align}\label{eq:PSD}
S_\text{IL}\left(\omega\right)=&\frac{2k_BT_{\textrm{0}}\mathit{\Gamma_{\rm 0}/m}}{{\left(\omega^2-{\omega_\text{z}}^2\right)}^2+{\mathit{\Gamma_{\rm t}}}^2\omega^2} \notag\\
&+\frac{{\left(\omega^2-{\omega_\text{z}}^2\right)}^2+\mathit{\Gamma_{\text{0}}}^2\omega^2}{{\left(\omega^2-{\omega_\text{z}}^2\right)}^2+{\mathit{\Gamma_{\rm t}}}^2\omega^2}S_{\textrm{nn}},
\end{align}}
where $\omega = 2\pi\times f$, $m$ is the particle's mass, $\omega_z$ is the mode frequency, $T_0$ is the bath temperature and $\mathit{\Gamma_{\rm t}} = \mathit{\Gamma_\text{0}} + \mathit{\Gamma_{\rm fb}}(\phi)$ is the total momentum damping rate on the particle's motion, with $\mathit{\Gamma_\text{0}}$ being the calculable momentum damping rate due to the pressure of the surrounding gas \cite{Millen2020rev} and $\mathit{\Gamma_{\rm fb}}(\phi)=\mathit{\Gamma_{\rm fb}}\textrm{cos}(\phi+\phi_0)$ being the additional damping controlled by the feedback gain, which is feedback-phase {$\phi$ dependent, with $\phi_0$ the uncontrollable phase delay caused by electronics and data processing}. The term $S_{\textrm{nn}}$ is the feedback circuit noise that is modeled {as white noise}. When $\mathit{\Gamma_{\rm fb}}$ is large the feedback can introduce extra noise and leads to heating, as discussed in \cite{Conangla2019,penny2021performance, melo2024vacuum}. The parameters $T_{\rm 0}$, $\mathit{\Gamma_{\rm 0}}$ and $\mathit{\Gamma_{\rm fb}}(\phi)$ can be extracted from a measured PSD by fitting equation~(\ref{eq:PSD}) to the data. Due to voltage noise from the amplifiers driving our Paul trap, the equilibrium temperature of our particles without cooling ranges from $T_0 = 400-1500\,\textrm{K}$, depending on their charge and spatial location in the trap, hence we express temperatures as a ratio. {The equilibrium temperatures for all particles in this paper are given in Supplementary Material S6.}

According to the equipartition theorem, the temperature of a levitated particle experiencing cold damping is \cite{Conangla2019,penny2023sympathetic, dania2021optical}: {
\begin{multline}\label{eq:Coolingphase}
T_{\text{CoM}}=\frac{\mathit{\Gamma_\text{0}}T_\text{0}}{\mathit{\Gamma_\text{0}}+\mathit{\Gamma_\textrm{{fb}}}\mathrm{cos}\left(\phi+\phi_0\right)}\\+\frac{m\omega_\text{z}^2{\mathit{\Gamma_{{\mathrm{fb}}}}^2{\textrm{cos}\left(\phi+\phi_\text{0}\right)}^2}}{2\text{k}_\text{B}\left(\mathit{\Gamma_\text{0}}+{\mathit{\Gamma_{\mathrm{fb}}}}\textrm{cos}\left( \phi+\phi_\text{0}\right) \right)}S_{\mathrm{nn}}.
\end{multline}
Experimentally, the temperature of each mode $T_{\text{CoM}}$ can be extracted from {equation~(\ref{eq:Coolingphase}) after fitting the measured PSD over the corresponding resonance peak with equation~(\ref{eq:PSD})} \cite{Conangla2019}.} In Fig.~\,\ref{fig:single-particle cooling}(b) we show the effect of increasing $\mathit{\Gamma_{\textrm{fb}}}$ on the temperature of a single mode of a single particle. 

The cooling depends on the phase between the feedback signal, which is proportional to velocity, and the detected motion of the particle. Equation~(\ref{eq:Coolingphase}) is fit to experimental data as $\phi$ is varied in Fig.~\ref{fig:single-particle cooling}(c), with $\mathit{\Gamma_\textrm{{fb}}}$ and $\phi_0$ as free parameters. The fitted value of $\phi_0$ is {$370^{\circ}\pm5^{\circ}$, noting that one period} of phase delay does not significantly effect cooling for an underdamped oscillator \cite{debiossac2020}, see Supplementary Materials S4.
The value of $\mathit{\Gamma_\textrm{{fb}}}$ obtained by fitting the data in Fig.~\ref{fig:single-particle cooling}(c) with equation~(\ref{eq:Coolingphase}) agrees with the value obtained at the same feedback gain by fitting the data with equation~(\ref{eq:PSD}), ($\mathit{\Gamma_\textrm{{fb}}}/(2\pi) = (0.82\pm0.05)\,$Hz, $(0.70\pm0.09)\,$Hz respectively. 

{Finally, in Fig.~\ref{fig:single-particle cooling}(d) we show the variation in temperature with $\mathit{\Gamma_\text{0}}$ by reducing the gas pressure, with $\mathit{\Gamma_\textrm{{fb}}}$ and $\phi$ fixed at around $6$\,Hz and $0^{\circ}$, respectively. At a pressure of $10^{-3}$ mbar, we reach the noise floor of our system, indicated by the grey region, at a temperature corresponding to $T_{\textrm{CoM}}=(6.8 \pm 0.4)\,\textrm{K}$, representing 17\,dB of cooling. The optimal $\mathit{\Gamma_\textrm{{fb}}}$ can be derived from equation \eqref{eq:Coolingphase}, see Supplementary Materials S7}. To further improve cooling one can improve particle illumination and imaging, decrease the noise in the levitation electronics, and replace the GTA of the EBC with an optimized tracking algorithm. Object tracking has the potential to track levitated microparticles at the shot-noise limit \cite{Werneck2024}. 

\subsubsection*{Simultaneous cooling of microparticles in an array}\label{subsec3}
Our neuromorphic imaging system {tracks} the motion of every object it identifies. We are able to process this information, and make a feedback loop for each degree-of-freedom that is detected. Each feedback loop consists of a dedicated FPGA and set of analogue filters, and we are limited in our experiment to three loops in total. We stress that this is not a limitation of detection or processing power, simply the number of FPGA outputs available to us. For each degree-of-freedom the phase and gain of each feedback loop must be optimized, and filters must be set accordingly.

\begin{figure}[t!]
\centering 
\includegraphics[width=0.475\textwidth]{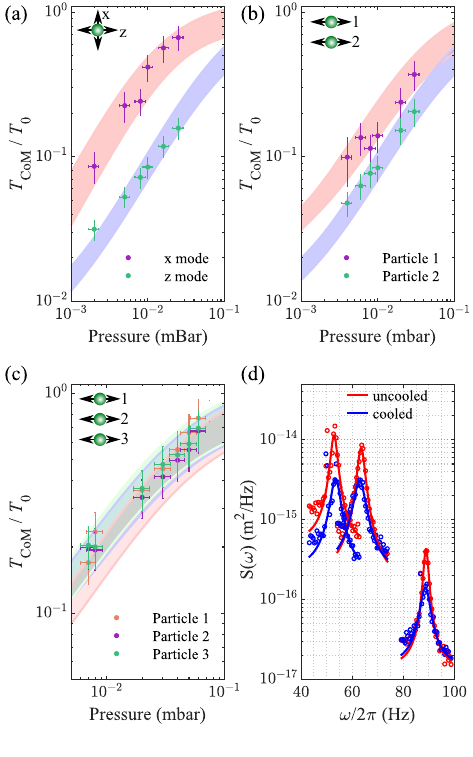}
\caption{\textbf{Simultaneous {multi-mode and multi-particle cooling.}} (a) Cooling of two orthogonal modes of a single particle's motion along the $x-$ and $z-$directions. (b) Cooling of two levitated particles' motion along the $z-$direction. (c) Cooling of three levitated particles' motion along the $z-$direction. Subfigures (a-c) also include the model from equation~(\ref{eq:Coolingphase}) as shaded coloured regions, with the area representing uncertainty in our experimental parameters. (d) The PSDs of the motion along the $z-$direction of the three particles in (c) before and after cooling (particle 1, 3, 2 in order of increasing frequency). Solid lines are fits to the model in equation~(\ref{eq:PSD}) which are used to extract the parameters used in the models (shaded regions) in (a-c). {All experimental error-bars in the figure are derived by taking 15 repeat experiments at each set of parameters to calculate a mean and standard deviation.}}
\label{fig:multi-particle cooling}
\end{figure}

In Fig.~\ref{fig:multi-particle cooling}(a) we cool two orthogonal degrees-of-freedom (the $x-$ and $z-$oscillation modes) of a single particle. We are sensitive to all three degrees-of-freedom due to the angle our imaging system makes to the principal axes of the trap, see Supplementary Methods S1. The geometry of our Paul trap allows the control of all degrees-of-freedom with a single electrode. 

We optimize and fix the feedback parameters, then lower the background pressure to reduce $\mathit{\Gamma_\text{0}}$, hence lowering the temperature $T_{\textrm{CoM}}$. The temperature is limited not by our noise floor, but by imperfect filtering pumping energy from the feedback signal into the $y-$mode, since it is close in frequency to the $z-$mode. At low pressures this causes the particle to become unstable, preventing us from further lowering the pressure. {Cross-talk between particle modes would also limit the ultimate cooling temperature, which could be resolved by better design of the feedback electrodes \cite{Gosling_2025}}.

In Fig.~\ref{fig:multi-particle cooling}(b) we extend our cooling to the $z-$mode of two separate particles. As we lower the pressure, the temperature of each mode drops, reaching $\sim10\,$dB and $\sim15\,$dB of cooling. Since only high-pass filters are used when cooling in the $z-$direction (see Methods), unfiltered noise from one particle is able to heat the uncooled modes of the other. At lower pressures this causes particle instability and prevents further cooling.

In Fig.\,\ref{fig:multi-particle cooling}(c) we cool the $z-$mode of three different particles, with the corresponding PSDs shown in Fig.\,\ref{fig:multi-particle cooling}(d). The issue of imperfect filtering is more pronounced when dealing with more particles, as there are more modes of the system overlapped with the unfiltered noise. We still achieve better than 7\,dB of cooling. Filtering can be improved either by separating the particle modes and applying band-pass filtering, or through the use of phase locked loops \cite{Jain2016, Vovrosh2017}. For the data in Figs.~\,\ref{fig:multi-particle cooling}(b-d) we adjust the particle spacing via the Paul trap voltages {until the Coulomb interaction is weak enough such that there are no coupled modes, see Supplementary Materials S2.} Collective modes in particle arrays can be cooled via sympathetic cooling {in the limit that the feedback damping rate is smaller than the coupling strength between the modes} \cite{penny2023sympathetic, bykov20233d}.

The EBC used in this study has a sensor size of $640\times 480$ pixels, with each particle image occupying $25\times25$ pixels (the coloured boxes in Fig.~\,\ref{fig:minion}(c)) and having a motional amplitude of 4 pixels. As long as the centre of the particles are separated by approximately 60 pixels, the particles can be individually tracked. Hence, without changing our imaging system we could simultaneously track of order 500 particles with this EBC. Considering the fact that object tracking allows for sub-pixel resolution \cite{Bullier2019, Ni2012}, by changing the magnification of the imaging system this EBC would be capable of simultaneously tracking at least 2000 levitated microparticles, with a correspondingly high data volume. 

\subsection*{Discussion}\label{sec4}
We have presented a scalable method for the detection and control of microparticles levitated in an array using a single neuromorphic detector. Neuromorphic imaging is ideally suited to this task, due to its natural affinity with detecting the motion of multiple objects \cite{Ni2015, Apps2025} and low data-transfer rate \cite{roy2019towards}. The tracking speed {in our work }is limited by the proprietary tracking algorithm of the EBC. {Commercial neuromorphic imaging sensors, such as the one used in this study, transfer data from the sensor to the camera hardware at GHz rates \cite{gallego2020event}}. The development of custom algorithms has enabled object tracking at 30\,kHz by working with the asynchronous data streamed from a DVS {using only 4MB of RAM on a standard 2.9 GHz Dual Core CPU} \cite{Ni2012}. {For a fixed frame rate the data volume is fixed regardless of the particle motion frequency. If the frame-rate is increased data is transferred more rapidly from the camera and the data volume and bandwidth increases.} By pushing above 100\,kHz, neuromorphic sensors would be suitable for feedback cooling optically levitated particles to the quantum ground state of motion \cite{magrini2021real}, considering the shot-noise limited potential of object tracking \cite{Bullier2019, Werneck2024}. This would require custom tracking algorithms {and interfacing the sensor directly with FPGA or neuromorphic processing electronics \cite{roy2019towards, Furmonas2022}, which would also enable the read-out and control of object alignment and rotation \cite{Kim2016, Apps2025}}.

We believe that the particle control method presented in this work could be extended to an array of order 100 microparticles, {(see Supplementary Materials S8)}. Multichannel FPGA systems with high-quality digital filters are a common tool in research labs. Paul traps are stable at low pressures \cite{bykov2019direct}, where the motional frequencies of levitated particles have sub-Hz linewidths \cite{pontin2020ultranarrow}. The naturally varying charge-to-mass ratio of charged microparticles, along with application of electric field gradients, will enable the spectral separation of motional modes, making possible single-particle control and cooling even for large arrays. The neuromorphic detection and cooling presented here is independent of the levitation method, as long as there is optical illumination, meaning it is suitable for small dielectrics in optical traps, charged absorptive materials in Paul traps (such as organic material) \cite{Conangla2019}, and magnetically levitated objects. 

Since the motion of levitated sensors is well understood, simple machine-learning could be used to optimize all of the feedback parameters \cite{Conangla2019} in an array, and optimal tracking algorithms used which have been shown to enable quantum-level control \cite{magrini2021real}. When combined with the low power-consumption of neuromorphic imaging technology (less than 30\,mW per tracked particle, see Supplementary Materials S5), and great progress in chip-scale particle levitation \cite{melo2024vacuum}, integrated devices containing arrays of quantum sensors are closer to being a reality.

\subsection*{Methods}\label{secA1}

\subsubsection*{Experimental setup}
\begin{figure}[tb]
\centering 
\includegraphics[width=0.475\textwidth]{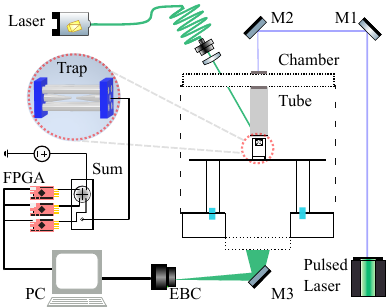}
\caption{{\bf Experimental setup for levitating, detecting and controlling arrays of microparticles.} A pulsed laser is used to launch particles into the trap via LIAD \cite{nikkhou2021}. Particles are illuminated with a CW laser from above, and imaged onto an EBC from below. The EBC software runs on a PC, and the tracking algorithm outputs data to a series of FPGAs, which process the data to produce a feedback signal for each degree of freedom of each particle. Each signal is then filtered with analogue filters (not shown), and then all signals are summed together and drive the control electrode.}
\label{fig:experimental setup}
\end{figure}

Figure~\ref{fig:experimental setup} shows the experimental setup surrounding our linear Paul trap. The trap consists of four parallel cylindrical trapping electrodes forming a square, with two coaxial cylindrical endcap electrodes, {one of which we call the control electrode which is used for feedback control}. The distance from trap centre to the surface of the $1\,\textrm{mm}$-diameter trapping electrodes is $r=1.15 \,\textrm{mm}$, and one opposing-pair are driven with 360\,V amplitude at 1\,kHz. {Of the other pair of the four electrodes, the lower one has a constant voltage of 3V applied to minimize single-particle micromotion.} The two endcaps are $300\,\mu\textrm{m}$ diameter and separated by $800\,\mu\textrm{m}$, {with a slight misalignment along the $z'-$axis which causes particles to be trapped along a diagonal in the $y'-z'$ plane}. The proximity of the electrodes to the centre of the trap means that a voltage applied to either one will create a field with a significant component in all three axes, enabling 3D control with a single electrode. 

Laser induced acoustic desorption (LIAD) \cite{nikkhou2021} is used to launch microparticles into the trap at a pressure of $1\,$mbar and then the pressure is lowered to carry out the experiments presented in this manuscript. We typically trap particles of positive charge, ranging from $2\times10^3\,\textrm{e}$ to $2\times10^4\,\textrm{e}$.

To image the particles, 18\,mW of laser light of wavelength $520\,$nm is weakly focused
onto the array. The scattered light is imaged onto an event-based camera (Prophesee EVK1 -Gen3.1 VGA (camera sensor: Prophesee PPS3MVCD, $640\times480$ pixels)) using a long working distance microscope. The camera is precisely calibrated using a method outlined in detail in \cite{ren2022event}. When dealing with multiple particles, calibration is performed via displacing a translation stage on which our imaging system is mounted by a known amount.

\subsubsection*{Data processing}
\begin{figure}[tb] 
\centering
\includegraphics[width=0.475\textwidth]{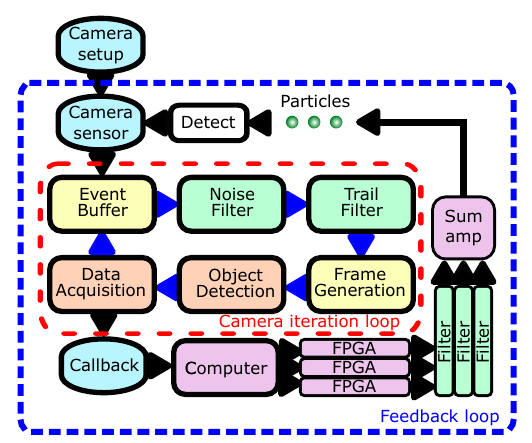}
\caption{{\bf Data pipeline for feedback control based on neuromorphic imaging.} {The EBC takes asynchronous data from the neuromorphic sensor, bundles it into events, accumulates the data over time to generate frames, and then detects and identifies objects. This information is passed to a computer, which uses a generic tracking algorithm (GTA) to track the motion of each object in 2D. The tracking data for each object is split into two 1D data-streams using simple Python code, and then each data stream is sent to a separate FPGA. The FPGAs each calculate the velocity from the position data, add a variable gain and phase shift, and generate voltage outputs. These are separately filtered using analogue filters, and the signals are combined with a summing amplifier, the output of which is sent to the control electrode to cool the particles.}}
\label{fig:pipeline of FPGA output}
\end{figure}

The data pipeline is shown in Fig.~\ref{fig:pipeline of FPGA output}. The operation of the EBC is described in detail in \cite{ren2022event}. The generic tracking algorithm (GTA) of the EBC outputs 2D position data for each object it detects. The EBC is communicated with via a Python script, which separates this 2D information into two 1D data streams for each object. The script passes each data stream to one of three FPGA systems (Red Pitaya STEMlab 125-14) to output the position of each particle. {The FPGA clock is synchronized to the clock of the EBC to ensuring timing consistency.} Code on the FPGA computes the velocity from the position data, and adds a variable gain and phase to the signal to generate the feedback signal. There is a latency of 10\,ms in this pipeline, {see Supplementary Materials S4}. EBCs are available with FPGA systems on the camera hardware, which will significantly reduce this latency.

Each feedback signal is filtered with an analogue filter to isolate each frequency component of motion: a high-pass filter for the $f_z$ signal (Wavefonix 3320 HPF 24\,dB per Octave) and a low-pass filter for the $f_x$ signal (Wavefonix 2140 LPF 24\,dB per Octave). The filtered feedback signals are combined with a summing amplifier, and sent to the control electrode.

\subsection*{Data Availability}
The data supporting this article is openly available from the King’s College London research data repository, KORDS, at https://doi.org/10.18742/28069136

\subsection*{Code Availability}
The code used in the present work is available from the corresponding
authors upon request.

\printbibliography

\subsection*{Acknowledgments}
We would like to thank Prof David Moore for useful discussions, Dr Ruvi Lecamwasam for assistance with writing FPGA code, and Dr John Dale for bringing neuromorphic imaging to our attention. This work is supported by STFC Grant ST/Y004914/1, EPSRC Grant EP/S004777/1 and ERC Starting Grant 803277.

\subsection*{Author Contributions}
YR - experimental design and build, data taking, data analysis. BS - experimental design. RY - data taking. QW - supporting simulations. JP - data analysis. MR - experimental design. JM - experimental design, data analysis. 

\subsection*{Competing Interests}
The authors declare no relevant competing interests.

\onecolumn
\newpage
\appendix
\begin{refsection}

\titleformat{\section}{\normalfont\Large\bfseries}{S\thesection}{1em}{}
\renewcommand{\thesection}{\arabic{section}}
\renewcommand{\figurename}{Supplementary Figure}
\renewcommand{\tablename}{Supplementary Table}
\crefname{table}{Supplementary Table}{Supplementary Tables}
\crefname{figure}{Supplementary Figure}{Supplementary Figures}
\setcounter{figure}{0}

\section*{\huge Supplementary Materials}

\baselineskip18pt

In this Supplemental Material we provide technical details on the experiment and data analysis, including: S1 Coordinate systems; S2 Mode identification for multiple particles; S3 Extended information for the four-particle dataset; S4 Latency in the feedback loop; S5 Power consumption of the event based camera; S6 Bath temperatures $T_0$ for all particles; S7 Cooling limit and the noise squashing; S8 Scalability of multi-mode cooling.

\section{Coordinate systems}
\begin{figure}[htbp!]
    \centering
    \includegraphics[width=0.5\linewidth]{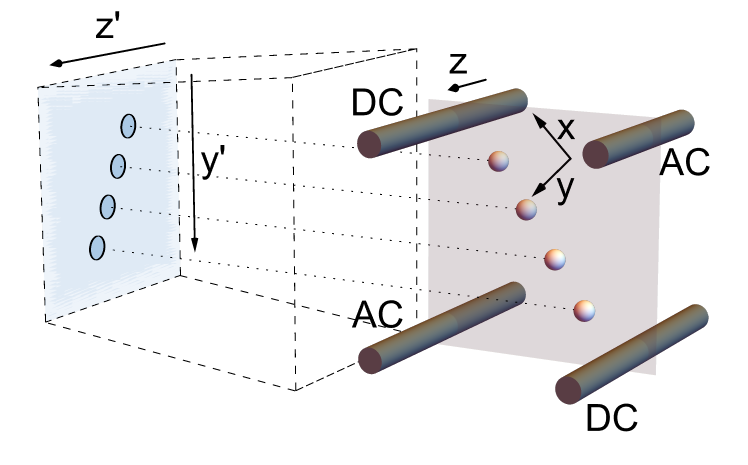}
    \caption{\textbf{Paul trap and camera coordinate systems.} The Paul trap coordinates defines the axes of oscillation of the levitated particles. The $x$- and $y$-axes are along the diagonals perpendicular to DC and AC electrodes, and the $z$-axis is parallel to the electrodes. The Paul trap $x$- and $y$-axes make a \ang{45} projection onto the camera $y'$-axis. The Paul trap $z$- axis is parallel to the camera $z'$-axis. An example of the projection of four trapped microparticles is illustrated [cf. Fig.~2 of the manuscript].}
    \label{fig:coordinate_systems}
\end{figure}

Our setup has two coordinate systems: the Paul trap coordinates $\{x,y,z\}$, and the camera coordinates $\{y',z'\}$, as shown in \cref{fig:coordinate_systems} and Fig.~1 of the manuscript. The Paul trap coordinates define the oscillation axes of the levitated particles. The $x$-axis is defined along the diagonal between the two trapping electrodes held at a DC voltage (see Methods in the main manuscript), the $y$-axis along the diagonal between the two trapping electrodes with an AC voltage, and the $z$-axis along the axis parallel to the endcap electrodes. In the camera frame, the trap $x$- and $y$-axes are projected onto the camera's $y'$-axis, and the camera's $z'$-axis is parallel to the $z$-axis. Therefore, we can capture the 3D motion of the levitated particles with the camera's 2D image. 

\newpage 

\section{Mode identification for multiple particles}\label{sec:mode_identifying}

\begin{figure}[tb]
    \centering
    \includegraphics{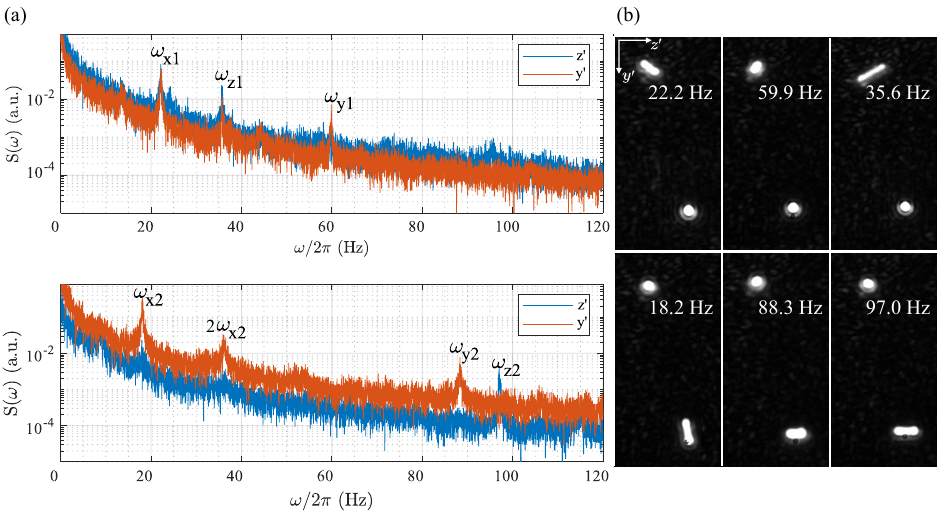}
    \caption{ \textbf {Mode identification for multiple levitated particles.} \textbf{(a)} PSDs of two levitated microparticles separated by 570\,$\mu$m. The upper (lower) figure shows the PSDs along the $y'$- and $z'$-axes for particle 1 (2) respectively. \textbf{(b)} A CMOS camera view as the particles are sequentially resonantly excited by a sinusoidal voltage at each of the six frequencies seen in (a).}
    \label{fig:two_particles_PSDs}
\end{figure}

Here we explain the method of identifying modes when we levitate multiple particles in an array, using a two-particle case study. We trap two charged microparticles separated by $570\,\mu$m, with the separation controlled by Paul trap voltages. Via neuromorphic detection, we calculate the power spectral densities (PSDs) of the two particles, as shown in \sfref{fig:two_particles_PSDs}{(a)}. We identify six modes in total. To distinguish these modes, we resonantly drive the Paul trap using a sinusoidal voltage applied to the control endcap electrode, at each of the six frequencies sequentially, and observe the response of the particles using a CMOS camera, as shown in \sfref{fig:two_particles_PSDs}{(b)}. For the upper particle the centre-of-mass motion frequencies are with $\omega_{x1}=22.2$\,Hz, $\omega_{y1}=59.9$\,Hz, and $\omega_{z1}=35.6$\,Hz and for the lower particle the centre-of-mass motion frequencies are $\omega_{x2}=18.2$\,Hz, $\omega_{y2}=88.3$\,Hz, and $\omega_{z2}=97.0$\,Hz.

\begin{figure}[htb!]
    \centering
    \includegraphics{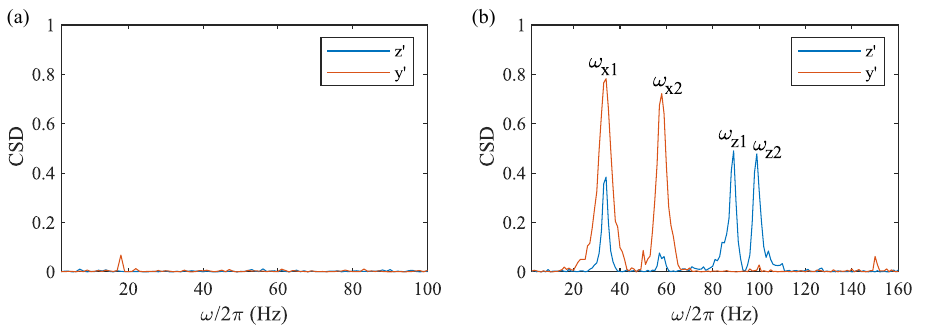}
    \caption{\textbf{CSDs of two levitated particles along to identify collective modes.} \textbf{(a)} CSDs for two particles separated by $570\,\mu$m, showing no coupled modes. \textbf{(b)} CSDs for two particles separated by $150\,\mu$m. Two modes along the $x$- and $z$-axes are coupled whilst no coupling is seen along the $y$-axis.}
    \label{fig:cross_spectral}
\end{figure}

To identify collective modes due to interactions between the charged particles, we calculate the cross-spectral density (CSD), which picks-out only the correlated spectral components \cite{penny2000signal}. As shown in \sfref{fig:cross_spectral}{(a)}, none of the six modes exhibit significant coupling, since the separation between the particles ($570\,\mu$m) is too large. 

When the particles are brought into close proximity, we do observe interactions between them. We trap another two particles and reduce the separation to approximately $150 \,\mu$m. We again observe six modes: 33.7 Hz and 58.1 Hz along the $x$-axis, 85.2 Hz and 150.9 Hz along the $y$-axis, and 89.4 Hz and 98.2 Hz along the $z$-axis. A CSD analysis in this case is given in \sfref{fig:cross_spectral}{(b)}, from which we can observe that both $x$-axis modes and both $z$-axis modes are correlated, whilst the $y$-axis modes appear to be uncoupled. This analysis is extended to the four particles in the manuscript below.

\newpage

\section{Extended information for the four-particle dataset}
\begin{figure}[htbp!]
    \centering
    \includegraphics[width=0.99\textwidth]{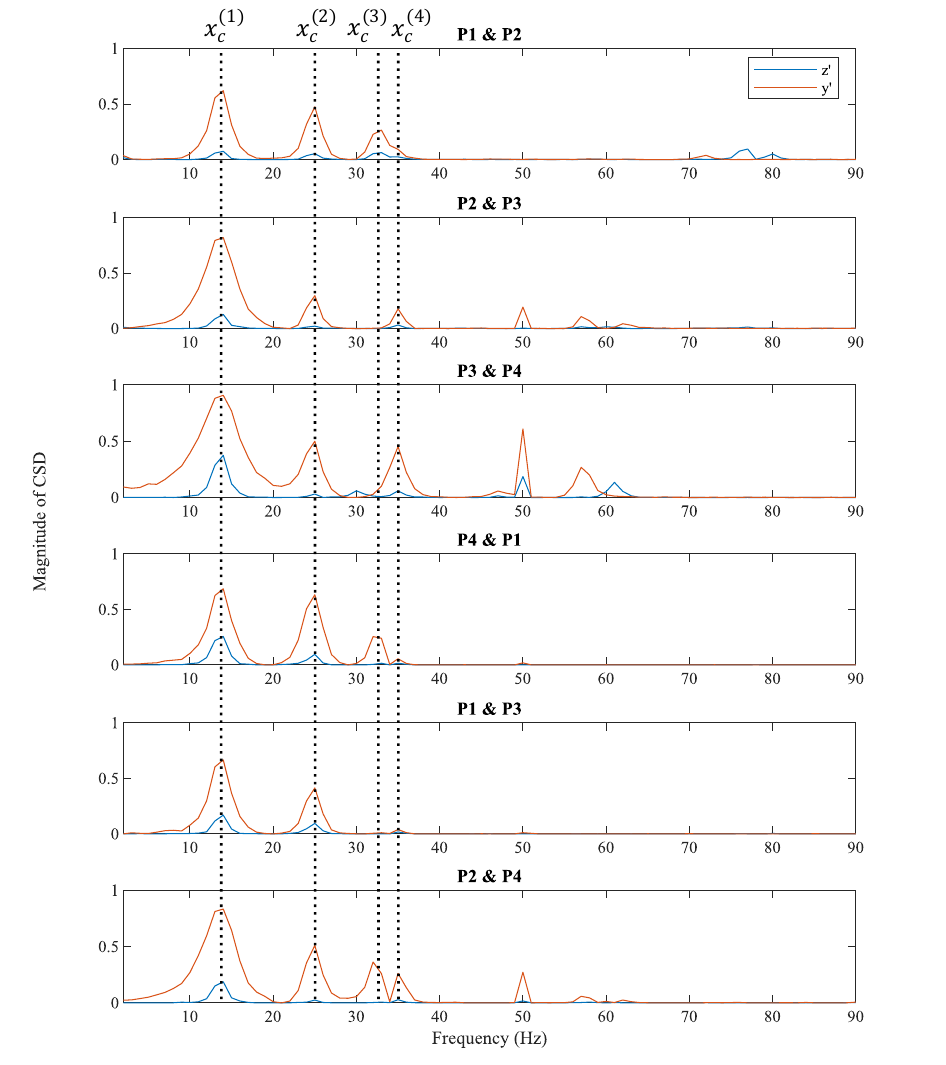}
    \caption{\textbf{CSD between the four levitated particles presented in Fig.~2 of the manuscript.} The four particles are labelled P1-4. Four coupled modes along the $x$- axis are observed, and we see no significant coupling between the $y$- and $z$-axis modes. }
    \label{fig:CSD_four_particles}
\end{figure}

Here we investigate the collective modes of the four levitated microparticles case shown in Fig.~2 of the manuscript. As discussed in the \hyperref[sec:mode_identifying]{Supplementary Section~S2}, the collective modes can be found via a CSD analysis. We show CSDs for all pairs of particles in \cref{fig:CSD_four_particles}, and conclude that the four levitated particles are coupled along the Paul trap $x$-axis (14\,Hz, 25\,Hz, 32\,Hz and 35\,Hz) due to their layout in the trap [cf.~\cref{fig:coordinate_systems}]. It is expected that four coupled oscillators have four coupled modes, but the nature of them is intricate and cannot be simply attributed as common or breathing modes unlike the two particle case \cite{penny2023sympathetic,Watson_2020}. 

In contrast, there is only evidence of very small correlations along the $y$- and $z$-axes, and only between nearest-neighbour particles. We do not see these modes in Fig.~2 of the manuscript, and if they are real they are below our detection noise-floor. Therefore, we conclude that the motion of the four levitated microparticles is only coupled along the $x$-axis.

We report the signal-to-noise ratios (SNRs) of our detection for each of the four particles in \cref{tab:4_particle_SNR}. We can see the strongest SNR is reported for particle P4, and it gradually reduces for particles further away from this. This is because we use a single laser for illumination focusing on P4. In single-particle experiments, we achieve a SNR up to $35$\,dB \cite{ren2022event}. The SNR can be improved in future experiments using a higher-power illumination system which can be expanded to uniformly illuminate many particles. Alternatively, separate beams could be used for each particle, via e.g. a spatial light modulator \cite{obata2010multi, rieser2022tunable}. {The charge to mass ratios of the four levitated particles are provided in \cref{tab:4_particle_qm}, from which we can observe that the four charges are different, since the masses are the same to within 10\% (manufacturer specified).} 

\begin{table}[htbp!]
\centering
\begin{tabularx}{0.8\textwidth}{ 
|>{\centering\arraybackslash}p{3cm}
|>{\centering\arraybackslash}X
|>{\centering\arraybackslash}X
|>{\centering\arraybackslash}X
|>{\centering\arraybackslash}X| }
    \hline
    & Particle 1 & Particle 2 & Particle 3 & Particle 4 \\
    \hline
    $f_z$ SNR (dB) & $3.9\pm0.8$ & $6.2\pm1.1$ & $7.4\pm1.1$ & $27.9\pm1.1$ \\
    \hline
    $f_y$ SNR (dB) & $2.9\pm1.1$ & $6.5\pm1.1$ & $6.1\pm1.1$ & $19.8\pm1.1$ \\
    \hline
    $x_c^{(1)}$ SNR (dB) & $4.5\pm0.8$ & $10.9\pm0.8$ & $13.3\pm0.8$ & $15.3\pm0.8$ \\
    \hline
    $x_c^{(2)}$ SNR (dB) & $6.0\pm1.1$ & $5.0\pm1.3$ & $4.7\pm1.0$ & $12.1\pm1.1$ \\
    \hline
    $x_c^{(3)}$ SNR (dB) & $3.2\pm0.8$ & $5.1\pm1.4$ & $0$ & $15.2\pm0.9$ \\
    \hline
    $x_c^{(4)}$ SNR (dB) & $0$ & $3.2\pm1.1$ & $3.7\pm1.0$ & $15.6\pm0.9$ \\
    \hline
\end{tabularx}
\caption{\textbf{Detected SNRs of the four levitated particles.}}
\label{tab:4_particle_SNR}
\end{table}

\begin{table}[htbp!]
\centering
\begin{tabularx}{0.8\textwidth}{ 
|>{\centering\arraybackslash}p{3cm}
|>{\centering\arraybackslash}X
|>{\centering\arraybackslash}X
|>{\centering\arraybackslash}X
|>{\centering\arraybackslash}X| }
    \hline
    & Particle 1 & Particle 2 & Particle 3 & Particle 4 \\
    \hline
    q/m (C/kg) & $(470 \pm 30) \times 10^{-5}$ & $(430 \pm 10) \times 10^{-5}$ & $(380 \pm 30) \times 10^{-5}$ & $(310 \pm 10) \times 10^{-5}$ \\
    \hline
\end{tabularx}
\caption{\textbf{Charge to mass ratios of the four levitated particles.}}
\label{tab:4_particle_qm}
\end{table}

\newpage

\section{Latency in the feedback loop}
\begin{figure}[tb]
    \centering
    \includegraphics{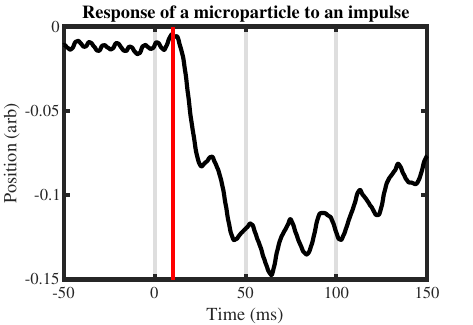}
    \caption{\textbf{Latency of the data pipeline.} A levitated particle is kicked with a voltage impulse applied to the control endcap electrode at time $t=0$. The response of the particle is recorded at the last step of the data pipeline (the output from the FPGA) and averaged over 40 realisations. The response is within 10ms, as indicated by the red line.}
    \label{fig:pipeline_latency}
\end{figure}

We measure the latency of the entire data pipeline by exposing a levitated particle to a voltage impulse and then monitoring the response in the data streamed from the last step in the pipeline (after the FPGA). In \cref{fig:pipeline_latency}, we present the result averaged over 40 realisations. We find that the response is within 10\,ms.

Our data-pipeline is sub-optimal, since data is transferred from the event based camera via a PC to the FPGA. In the future we aim to use an FPGA directly connected to the camera hardware. In such a case, the primary limitation would be the sensor latency ($200\,\mu$s) and the communication bandwidth between the camera hardware and a multi-channel DAC. However, in our current system: 
\begin{itemize}
    \item Latency from the changes in light intensity at each pixel to event output: typical 200\,$\mu$s from the detector manual.
    \item Data Transfer to Computer: The camera uses USB 3.0 at a maximum data-rate of 4 GBit/s – this will never be the limiting case because it’s much larger than the data transfer at maximum event-rate. USB 3.0 has a latency of about 30\,$\mu$s. 
    \item The EBC software then tracks the objects – latency in this step is very hard to evaluate, since the process is proprietary.
    \item Python code processes the tracking data for it to be sent to the FPGA, this is simple and can be considered negligible. 
    \item The data is transferred via Ethernet and a network switch to our collection of FPGAs, which has a latency of about 300\,$\mu$s. 
    \item Data Processing on FPGA: Each FPGA computes velocity, applies gain and a phase shift, and outputs a feedback signal. Since our FPGA runs at 124 Msps, it is reasonable to say the latency is negligible compared to the data transfer delays.
\end{itemize}

The delay of applied control signal has two effects (following Ref.~\cite{debiossac2020}). Firstly, it deterministically shifts the phase between the mechanical system and the feedback signal, leading to oscillatory behaviour in the correlations and thus a transition between heating and cooling. Secondly, it causes stochastic dephasing of the mechanical motion relative to the feedback signal as the delay becomes large, which reduces the strength of the correlations. 

From the experimental results of Ref.~\cite{debiossac2020}, correlation functions of the microparticle centre-of-mass motion as a function of delay remain high for short delays (on the order of a few oscillation periods) when the dynamics are underdamped (as in our system).
Since our levitated oscillators have oscillation frequencies below 100\,Hz, the latency of 10\,ms corresponds to a single period of delay, and has little consequence for our cooling protocol.

\section{Power consumption of the event based camera}
From the neuromorphic sensor manual (PPS3MVCD) the static power consumption of the EBC is 26\,mW, plus a dynamic power consumption based on sensor activity. In our experiments the event rate is about 500\,kevt/s, leading to a power consumption of about 27.6\,mW. This is very low compared to a CMOS camera (Thorlabs CS165MU/M, 1.17\,W Max @ 34.8 fps Full Sensor ROI) or a high-speed camera (iX Cameras i-SPEED 230, 17\,W at 2500\,fps). 

\section{Bath temperatures \texorpdfstring{$T_0$}{T0} for all particles}
Due to voltage noise from the amplifiers driving our Paul trap, the equilibrium temperature of our particles without cooling ranges from approximately $T_0 = 400-1500\,$K. This temperature varies depending on their charge and spatial location within the trap. Here we model the electrical noise as a white noise bath and assume the particles are trapped in a quadratic potential. Under the assumption that the equipartition theorem holds, the different bath temperatures used in the main text figures are listed in \cref{tab:T0_particles}. In Fig. 4(c), P2 is located closest to the trap center and therefore experiences the least electric field noise, resulting in the lowest bath temperature. In contrast, P1 and P3 are farther from the trap center, and their bath temperatures are approximately an order of magnitude higher than that of P2. A similar trend can be observed in Fig. 4(b), where P1, being farther from the trap center, also exhibits a higher temperature than P2. 
\begin{table}[htbp!]
\centering
\begin{tabularx}{0.905\textwidth}{ 
|>{\centering\arraybackslash}p{1.1cm}
|>{\centering\arraybackslash}p{1.5cm}
|>{\centering\arraybackslash}p{1.6cm}
|>{\centering\arraybackslash}p{1.5cm}
|>{\centering\arraybackslash}p{1.5cm}
|>{\centering\arraybackslash}p{2.1cm}
|>{\centering\arraybackslash}p{2.1cm}| }
    \hline
    & Fig3(b) & Fig3(c) & Fig3(d) & Fig4(a) & Fig4(b) & Fig4(c) \\
    \hline
    $T_0$ (K) & $900 \pm 300$ & $1500 \pm 100$ & $400 \pm 50$ & $500 \pm 70$ & P1:$2300 \pm 300$\newline P2:$600 \pm 100 $& P1:$3300 \pm 600$\newline{P2:$400 \pm 50 $}\newline P3:$4300 \pm 800$ \\
    \hline
\end{tabularx}
\caption{\textbf{Different equilibrium temperature $T_0$ values in the main text figures.}}
\label{tab:T0_particles}
\end{table}

\section{Cooling limit and the noise squashing}

\begin{figure}[htb]
    \centering
    \includegraphics[width=0.75\linewidth]{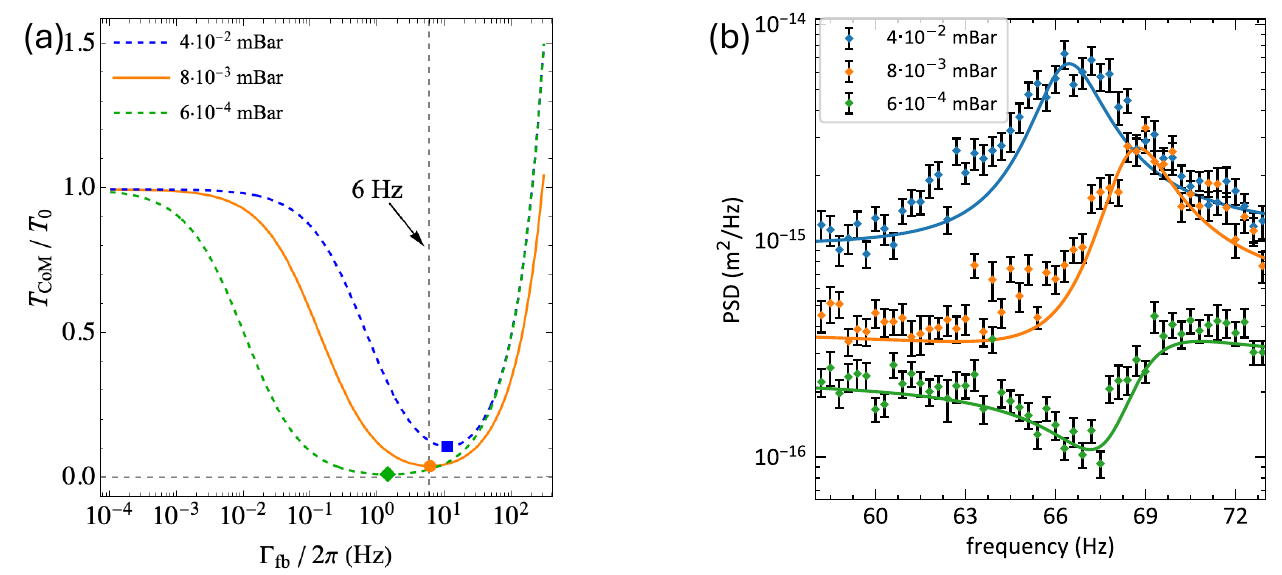}
    \caption{\textbf{Cooling limit and the noise squashing.} \textbf{(a)} The cooled temperature of the system $T_\textrm{CoM}$ against the feedback rate $\mathit{\Gamma_\textrm{fb}}$ at different pressures given by equation (2) in the main manuscript. Here the dots are the optimal feedback rates to reach the corresponding minimum cooling temperature, computed from \sfequ{equ:optimal_temperature}{Supplementary Equations (1)} and \sfequ{equ:optimal_feedback_rate}{(2)}. \textbf{(b)} The experimental PSDs of particle's measured motion fitted by \sfequ{equ:PSD_with_noise_squashing}{Supplementary Equations (3)}, showing the change from unsquashed (blue) to squashed (green) as the pressure reduces. Here we set feedback rate $\mathit{\Gamma_\textrm{fb}}/2\pi\approx6$ Hz.}
    \label{fig:cooling_and_squashing}
\end{figure}

At a certain pressure (which determines the gas damping rate $\mathit{\Gamma_\text{0}}$), to achieve maximal cooling the optimal feedback rate $\mathit{\Gamma_\textrm{fb}}$ can be obtained from equation~(2) in the main manuscript by taking $\partial T_\textrm{CoM}/\partial \mathit{\Gamma_\textrm{fb}} = 0$ \cite{melo2024vacuum}, which gives 
\begin{equation}\label{equ:optimal_temperature}
    T_\textrm{CoM}^\textrm{opt}=\frac{-m\mathit{\Gamma_\text{0}}\omega_z^2 S_\textrm{nn} + \sqrt{m\mathit{\Gamma_\text{0}}\omega_z^2S_\textrm{nn} (2k_BT_0 + m\mathit{\Gamma_\text{0}}\omega_z^2 S_\textrm{nn})}}{k_B}
\end{equation}
with the optimal feedback rate
\begin{equation}\label{equ:optimal_feedback_rate}
    \mathit{\Gamma_\textrm{fb}}^\textrm{opt}=\frac{1}{\cos(\phi)}\left(-\mathit{\Gamma_\text{0}} + \frac{ \sqrt{m\mathit{\Gamma_\text{0}}\omega_z^2S_\textrm{nn} (2k_BT_0 + m\mathit{\Gamma_\text{0}}\omega_z^2 S_\textrm{nn})}}{m \omega_z^2 S_\textrm{nn}}\right)
\end{equation}
Note that $T_\textrm{CoM}^\textrm{opt}$ is $\phi$-independent. 

The relation between the feedback rate $\Gamma_\textrm{fb}$ and the cooling temperature $T_\textrm{CoM}$ is shown in \sfref{fig:cooling_and_squashing}{(a)}, where we plot equation (2) in the main manuscript at three different pressures. The dots in the plot are the optimal feedback rates and the optimal temperatures corresponding to the respective pressure, computed from \sfequ{equ:optimal_temperature}{Supplementary Equations (1)} and \sfequ{equ:optimal_feedback_rate}{(2)}. As we can see, the optimal feedback rate gets smaller as the pressure goes down. 

At a certain pressure, a larger feedback rate $\mathit{\Gamma_\textrm{fb}}$ results in a lower system temperature $T_\textrm{CoM}$ (hence a better cooling), until it pass the the optimal value in which case the feedback signal instead heats up the system. This is because the increasing amplitude of the feedback signal also amplifies the noise fed back to the particle, preventing further cooling. The corresponding PSD also appears inverted below the noise floor, a phenomenon known as the noise squashing \cite{Poggio2007}. In \sfref{fig:cooling_and_squashing}{(b)} we demonstrate this by setting the feedback rate $\mathit{\Gamma_\textrm{fb}}/2\pi=6$ Hz and reducing the pressure. Correspondingly, the PSDs of the measured motion changes from unsquashed (blue line, $\mathit{\Gamma_\textrm{fb}}<\mathit{\Gamma_\textrm{fb}}^\textrm{opt}$), to mixed (orange line, $\mathit{\Gamma_\textrm{fb}}\approx \mathit{\Gamma_\textrm{fb}}^\textrm{opt}$), and finally to squashed (green line, $\mathit{\Gamma_\textrm{fb}}>\mathit{\Gamma_\textrm{fb}}^\textrm{opt}$).
Furthermore, the squashed shape is asymmetrical due to the delay in the signal \cite{Conangla2019}. This makes a small change to the fitting PSD function, as shown in the following:

Suppose we consider a time delay in the feedback signal $v(t) \to v(t-\tau)$, the equation of motion reads
\begin{equation*}
\overset{\mathrm{..}}{z}+\Gamma_g\overset{\mathrm{.}}{z}+\omega_0^2z=\frac{\sigma}{m}\eta(t)-\mathit{\Gamma_{fb}}(v(t - \tau)+\overset{\mathrm{.}}{\xi}(t)).
\end{equation*}
Take the Fourier transform with $z(\omega) = \mathcal{F}[z(t)]= \int_{-\infty}^\infty z(t)e^{-i\omega t}\mathrm{d}t$ and $\mathcal{F}[z(t-\tau)]=e^{-i\omega \tau} \mathcal{F}[z(t)]$, we get the particle's detected motion in the frequency domain
\begin{align*}
\chi'(\omega) + F(\xi(t))=&\frac{\sigma/m}{(\omega_0^2-\omega^2+\omega\mathit{\Gamma_\text{fb}}\sin(\omega\tau ))+i\omega(\mathit{\Gamma_\text{g}}+\cos(\omega\tau)\mathit{\Gamma_{\mathrm{fb}}})}F(\eta(t)) \notag\\
&\qquad+\frac{\omega_0^2-\omega^2+i\omega\mathit{\Gamma_\mathrm{g}} + i\omega\mathit{\Gamma_\text{fb}} (e^{-i\omega \tau}-1 )}{(\omega_0^2-\omega^2+\omega\mathit{\Gamma_\text{fb}}\sin(\omega\tau ))+i\omega(\mathit{\Gamma_\text{g}}+\cos(\omega\tau)\mathit{\Gamma_{\mathrm{fb}})}}F(\xi(t)),
\end{align*}
We can calculate the PSD in this situation by taking $S_z(\omega)=\mathrm{IE}(\lvert \chi'(\omega)^2\rvert)$ and assume that two noises $F(\eta(t))$ and $F(\xi(t))$ are uncorrelated, which gives the PSD of the measured position in the in-loop detector:
\begin{align}\label{equ:PSD_with_noise_squashing}
S_{\rm{IL}}(\omega)=&\frac{{\sigma^2}/{m^2}}{(\omega_0^2-\omega^2 + \omega\mathit{\Gamma_\text{fb}}\sin(\omega\tau))^2+(\mathit{\Gamma_\text{g}}+\cos(\omega\tau)\mathit{\Gamma_\text{fb}})^2\omega^2}\notag \\
&\qquad+\frac{(\omega_0^2-\omega^2 + \omega\mathit{\Gamma_\text{fb}}\sin(\omega\tau))^2+\omega^2 (\mathit{\Gamma_{\mathrm{g}}} + \mathit{\Gamma_\text{fb}}(\cos(\omega\tau)-1))^2}{(\omega_0^2-\omega^2 + \omega\mathit{\Gamma_\text{fb}}\sin(\omega\tau))^2+(\mathit{\Gamma_\text{g}}+\cos(\omega\tau)\mathit{\Gamma_\text{fb}})^2\omega^2}S_{nn}.
\end{align}
This signal delay $\tau$ can cause a small asymmetry that is visible in the squashed PSD from
in-loop detection. In this case, the experimental PSD of particle's measured motion can be correctly fitted by \sfequ{equ:PSD_with_noise_squashing}{Supplementary Equations (3)}, as shown in \sfref{fig:cooling_and_squashing}{(b)}.

\section{Scalability of multi-mode cooling}
Here we discuss the fundamental limit on the number of modes $N$ we can cool with the technique presented in our manuscript. We assume $N$ is bounded by the bandwidth of our detection $F\approx 500$\,Hz and the minimum resolvable linewidth $\Delta f$, such that $N \approx \frac{F}{\Delta f}$. Then, to estimate the number of modes, we need to find the minimum linewidth $\Delta f$.

The contribution to this limit from the experimental setup comes mainly from filtering. Considering the time duration for signal demodulation, there is a bandwidth limit for filters to pick out certain frequency signal which will set a limit to $\Delta f$. A PLL has a very narrow filter bandwidth, for example a Zurich Instruments HF2LI has a minimum bandwidth of $83\,\mu$Hz and can filter 6 peaks at once. Therefore, with PLL it is possible to pick up feedback signal at mHz resolution.

On the other hand, since we use cold damping to cool the particles, the particles inevitably experience an additional damping rate, which broadens the linewidth and makes the modes harder to distinguish. According to Eq. (2) in the manuscript, by solving $\partial T_\textrm{CoM}/\partial \mathit{\Gamma_\textrm{fb}}=0$, we can obtain an optimal $\mathit{\Gamma_\textrm{fb}}$ at a given pressure (See \sfequ{equ:optimal_feedback_rate}{Supplementary Equations (2)}). This optimal $\mathit{\Gamma_\textrm{fb}}$ would be the dominant factor in determining the scaling number, whose value reduces as pressure goes down. 

If we take the above argument as a rule of thumb, we can estimate the number of modes as follows. Suppose we start cooling at $10^{-4}$\,mbar in experiment, at which the particle’s linewidth is about $\mathit{\Gamma_\textrm{0}}/2\pi\approx0.002$\,Hz, the optimal feedback rate is $\mathit{\Gamma_\textrm{fb}}/2\pi\approx1.8$\,Hz. This limits the capacity of our setup to be $N\approx270$ modes, which can be improved at high vacuum.

\printbibliography[title = References for Supplementary Materials]

@article{Ni2015,
    author = {Ni, Zhenjiang and Ieng, Sio-Hoi and Posch, Christoph and Régnier, Stéphane and Benosman, Ryad},
    title = {Visual Tracking Using Neuromorphic Asynchronous Event-Based Cameras},
    journal = {Neural Computation},
    volume = {27},
    number = {4},
    pages = {925-953},
    year = {2015},
    month = {04},
    issn = {0899-7667},
    doi = {10.1162/NECO_a_00720},
    url = {https://doi.org/10.1162/NECO\_a\_00720},
    eprint = {https://direct.mit.edu/neco/article-pdf/27/4/925/934400/neco\_a\_00720.pdf}
}

@ARTICLE{Gosling_2025,
       author = {{Gosling}, J.~M.~H. and {Pontin}, A. and {Alder}, F. and {Rademacher}, M. and {Monteiro}, T.~S. and {Barker}, P.~F.},
        title = "{Feedback cooling scheme for an optically levitated oscillator with controlled cross-talk}",
      journal = {arXiv e-prints},
         year = 2025,
        month = 6,
          eid = {arXiv:2506.17172},
        pages = {arXiv:2506.17172},
          doi = {10.48550/arXiv.2506.17172},
archivePrefix = {arXiv},
       eprint = {2506.17172},
 primaryClass = {physics.optics},
       adsurl = {https://ui.adsabs.harvard.edu/abs/2025arXiv250617172G}
}

@article{Ni2012,
author = {NI, Z. and PACORET, C. and BENOSMAN, R. and IENG, S. and RÉGNIER*, S.},
title = {Asynchronous event-based high speed vision for microparticle tracking},
journal = {Journal of Microscopy},
volume = {245},
number = {3},
pages = {236-244},
doi = {https://doi.org/10.1111/j.1365-2818.2011.03565.x},
url = {https://onlinelibrary.wiley.com/doi/abs/10.1111/j.1365-2818.2011.03565.x},
eprint = {https://onlinelibrary.wiley.com/doi/pdf/10.1111/j.1365-2818.2011.03565.x},
year = {2012}
}

@article{Chircov2022,
  title     = "Microelectromechanical systems ({MEMS}) for biomedical
               applications",
  author    = "Chircov, Cristina and Grumezescu, Alexandru Mihai",
  journal   = "Micromachines",
  publisher = "MDPI AG",
  volume    =  13,
  number    =  2,
  pages     = "164",
  month     =  1,
  year      =  2022,
  copyright = "https://creativecommons.org/licenses/by/4.0/"
}

@article{Imboden2014,
title = {Dissipation in nanoelectromechanical systems},
journal = {Phys. Rep.},
volume = {534},
number = {3},
pages = {89-146},
year = {2014},
note = {Dissipation in nano-electromechanical systems},
issn = {0370-1573},
doi = {https://doi.org/10.1016/j.physrep.2013.09.003},
url = {https://www.sciencedirect.com/science/article/pii/S0370157313003475},
author = {Matthias Imboden and Pritiraj Mohanty}
}

@article{Shaeffer2013,
  author={Shaeffer, Derek K.},
  journal={IEEE Commun. Mag.}, 
  title={MEMS inertial sensors: A tutorial overview}, 
  year={2013},
  volume={51},
  number={4},
  pages={100-109},
  doi={10.1109/MCOM.2013.6495768}
}

@article{lichtsteiner2008128,
  title={A $128 \times 128$ $120\,$dB $15\,\mu$ s latency asynchronous temporal contrast vision sensor},
  author={Lichtsteiner, Patrick and Posch and Christoph and Delbruck, Tobi},
  journal={IEEE J. Solid-State Circuits.},
  volume={43},
  number={2},
  pages={566--576},
  year={2008},
  publisher={IEEE}
}

@article{mangalwedhekar2023achieving,
  title={Achieving nanoscale precision using neuromorphic localization microscopy},
  author={Mangalwedhekar, Rohit and Singh, Nivedita and Thakur, Chetan Singh and Seelamantula, Chandra Sekhar and Jose, Mini and Nair, Deepak},
  journal={Nat. Nanotechnol.},
  volume={18},
  number={4},
  pages={380--389},
  year={2023},
  publisher={Nature Publishing Group UK London}
}

@article{roy2019towards,
  title={Towards spike-based machine intelligence with neuromorphic computing},
  author={Roy, Kaushik and Jaiswal and Akhilesh and Panda, Priyadarshini},
  journal={Nature},
  volume={575},
  number={7784},
  pages={607--617},
  year={2019},
  publisher={Nature Publishing Group UK London}
}

@article{gehrig2024low,
  title={Low-latency automotive vision with event cameras},
  author={Gehrig, Daniel and Scaramuzza, Davide},
  journal={Nature},
  volume={629},
  number={8014},
  pages={1034--1040},
  year={2024},
  publisher={Nature Publishing Group UK London}
}

@article{gallego2020event,
  title={Event-based vision: A survey},
  author={Gallego, Guillermo and Delbr{\"u}ck, Tobi and Orchard, Garrick and Bartolozzi, Chiara and Taba, Brian and Censi, Andrea and Leutenegger, Stefan and Davison, Andrew J and Conradt, J{\"o}rg and Daniilidis, Kostas and others},
  journal={IEEE Trans. Pattern Anal. Mach. Intell.},
  volume={44},
  number={1},
  pages={154--180},
  year={2020},
  publisher={IEEE}
}

@inproceedings{nguyen2019real,
  title={Real-time 6DOF pose relocalization for event cameras with stacked spatial LSTM networks},
  author={Nguyen, Anh and Do, Thanh-Toan and Caldwell, Darwin G and Tsagarakis, Nikos G},
  booktitle={Proceedings of the IEEE/CVF Conference on Computer Vision and Pattern Recognition Workshops},
  year={2019}
}

@article{rebecq2019high,
  title={High speed and high dynamic range video with an event camera},
  author={Rebecq, Henri and Ranftl, Ren{\'e} and Koltun, Vladlen and Scaramuzza, Davide},
  journal={IEEE Trans. Pattern Anal. Mach. Intell.},
  volume={43},
  number={6},
  pages={1964--1980},
  year={2019},
  publisher={IEEE}
}

@inproceedings{cannici2019asynchronous,
  title={Asynchronous convolutional networks for object detection in neuromorphic cameras},
  author={Cannici, Marco and Ciccone, Marco and Romanoni, Andrea and Matteucci, Matteo},
  booktitle={Proceedings of the IEEE/CVF Conference on Computer Vision and Pattern Recognition Workshops},
  year={2019}
}

@article{zhu2023neuromorphic,
  title={Neuromorphic sequence learning with an event camera on routes through vegetation},
  author={Zhu, Le and Mangan and Michael and Webb, Barbara},
  journal={Sci. Robot.},
  volume={8},
  number={82},
  pages={eadg3679},
  year={2023},
  publisher={American Association for the Advancement of Science}
}

@article{dania2021optical,
  title={Optical and electrical feedback cooling of a silica nanoparticle levitated in a Paul trap},
  author={Dania, L. and Bykov, D. S. and Knoll, M. and Mestres, P. and Northup, T. E.},
  journal={Phys. Rev. Res.},
  volume={3},
  number={1},
  pages={013018},
  year={2021},
  publisher={APS}
}

@article{nikkhou2021,
AUTHOR = {Nikkhou, Maryam and Hu, Yanhui and Sabin, James A. and Millen, James},
TITLE = {Direct and Clean Loading of Nanoparticles into Optical Traps at Millibar Pressures},
JOURNAL = {Photonics},
VOLUME = {8},
YEAR = {2021},
NUMBER = {11},
pages={458},
url = {https://www.mdpi.com/2304-6732/8/11/458},
ISSN = {2304-6732},
doi = {10.3390/photonics8110458}
}

@article{debiossac2020,
author={Debiossac, Maxime
and Grass, David
and Alonso, Jose Joaquin
and Lutz, Eric
and Kiesel, Nikolai},
title={Thermodynamics of continuous non-Markovian feedback control},
journal={Nat. Commun.},
year={2020},
month={3},
day={13},
volume={11},
number={1},
pages={1360},
issn={2041-1723},
doi={10.1038/s41467-020-15148-5},
url={https://doi.org/10.1038/s41467-020-15148-5}
}

@article{melo2024vacuum,
author={Melo, Bruno
and T. Cuairan, Marc
and Tomassi, Gr{\'e}goire F. M.
and Meyer, Nadine
and Quidant, Romain},
title={Vacuum levitation and motion control on chip},
journal={Nat. Nanotechnol.},
volume={19},
year={2024},
month={6},
day={06},
issn={1748-3395},
pages={1270–1276},
doi={10.1038/s41565-024-01677-3},
url={https://doi.org/10.1038/s41565-024-01677-3}
}

@article{dania2024ultrahigh,
  title={Ultrahigh Quality Factor of a Levitated Nanomechanical Oscillator},
  author={Dania, Lorenzo and Bykov, Dmitry S and Goschin, Florian and Teller, Markus and Kassid, Abderrahmane and Northup, Tracy E},
  journal={Phys. Rev. Lett.},
  volume={132},
  number={13},
  pages={133602},
  year={2024},
  publisher={APS}
}

@article{afek2022coherent,
  title={Coherent scattering of low mass dark matter from optically trapped sensors},
  author={Afek Gadi and Carney and Daniel and Moore, David C},
  journal={Phys. Rev. Lett.},
  volume={128},
  number={10},
  pages={101301},
  year={2022},
  publisher={APS}
}

@article{yin2022experiments,
  title={Experiments with levitated force sensor challenge theories of dark energy},
  author={Yin, Peiran and Li, Rui and Yin, Chengjiang and Xu, Xiangyu and Bian, Xiang and Xie, Han and Duan, Chang-Kui and Huang, Pu and He, Jian-hua and Du, Jiangfeng},
  journal={Nat. Phys.},
  volume={18},
  number={10},
  pages={1181--1185},
  year={2022},
  publisher={Nature Publishing Group UK London}
}

@article{ahn2020ultrasensitive,
  title={Ultrasensitive torque detection with an optically levitated nanorotor},
  author={Ahn, Jonghoon and Xu, Zhujing and Bang, Jaehoon and Ju, Peng and Gao, Xingyu and Li, Tongcang},
  journal={Nat. Nanotechnol.},
  volume={15},
  number={2},
  pages={89--93},
  year={2020},
  publisher={Nature Publishing Group UK London}
}

@article{vijayan2023scalable,
  title={Scalable all-optical cold damping of levitated nanoparticles},
  author={Vijayan, Jayadev and Zhang, Zhao and Piotrowski, Johannes and Windey, Dominik and van der Laan, Fons and Frimmer, Martin and Novotny, Lukas},
  journal={Nat. Nanotechnol.},
  volume={18},
  number={1},
  pages={49--54},
  year={2023},
  publisher={Nature Publishing Group UK London}
}

@article{penny2023sympathetic,
  title={Sympathetic cooling and squeezing of two colevitated nanoparticles},
  author={Penny, TW and Pontin, A and Barker, PF},
  journal={Physical Review Research},
  volume= {5},
  number={1},
  pages={013070},
  year={2023},
  publisher={APS}
}

@article{bykov20233d,
  title={3D sympathetic cooling and detection of levitated nanoparticles},
  author={Bykov, Dmitry S and Dania, Lorenzo and Goschin, Florian and Northup, Tracy E},
  journal={Optica},
  volume={10},
  number={4},
  pages={438--442},
  year={2023},
  publisher={Optica Publishing Group}
}

@article{delic2020cooling,
  title={Cooling of a levitated nanoparticle to the motional quantum ground state},
  author={Deli{\'c}, Uro{\v{s}} and Reisenbauer, Manuel and Dare, Kahan and Grass, David and Vuleti{\'c}, Vladan and Kiesel, Nikolai and Aspelmeyer, Markus},
  journal={Science},
  volume={367},
  number={6480},
  pages={892--895},
  year={2020},
  publisher={American Association for the Advancement of Science}
}

@article{millen2020quantum,
  title={Quantum experiments with microscale particles},
  author={Millen, James and Stickler, Benjamin A},
  journal={Contemp. Phys.},
  volume={61},
  number={3},
  pages={155--168},
  year={2020},
  publisher={Taylor \& Francis}
}

@article{rudolph2022force,
  title={Force-gradient sensing and entanglement via feedback cooling of interacting nanoparticles},
  author={Rudolph, Henning and Deli{\'c}, Uro{\v{s}} and Aspelmeyer, Markus and Hornberger, Klaus and Stickler, Benjamin A},
  journal={Phys. Rev. Lett.},
  volume={129},
  number={19},
  pages={193602},
  year={2022},
  publisher={APS}
}

@article{Slezak2019,
    author = {Slezak, Bradley R. and D'Urso, Brian},
    title = {A microsphere molecule: The interaction of two charged microspheres in a magneto-gravitational trap},
    journal = {Appl. Phys. Lett.},
    volume = {114},
    number = {24},
    pages = {244102},
    year = {2019},
    month = {06},
	issn = {0003-6951},
    doi = {10.1063/1.5097615},
    url = {https://doi.org/10.1063/1.5097615},
    eprint = {https://pubs.aip.org/aip/apl/article-pdf/doi/10.1063/1.5097615/19774174/244102\_1\_online.pdf}
}

@article{Svak2021,
author = {Vojt\v{e}ch Svak and Jana Flaj\v{s}manov\'{a} and Luk\'{a}\v{s} Chv\'{a}tal and Martin \v{S}iler and Alexandr Jon\'{a}\v{s} and Jan Je\v{z}ek and Stephen H. Simpson and Pavel Zem\'{a}nek and Oto Brzobohat\'{y}},
journal = {Optica},
number = {2},
pages = {220--229},
publisher = {Optica Publishing Group},
title = {Stochastic dynamics of optically bound matter levitated in vacuum},
volume = {8},
month = {2},
year = {2021},
url = {https://opg.optica.org/optica/abstract.cfm?URI=optica-8-2-220},
doi = {10.1364/OPTICA.404851}
}

@article{Millen2020rev,
doi = {10.1088/1361-6633/ab6100},
url = {https://dx.doi.org/10.1088/1361-6633/ab6100},
year = {2020},
month = {1},
publisher = {IOP Publishing},
volume = {83},
number = {2},
pages = {026401},
author = {James Millen and Tania S Monteiro and Robert Pettit and A Nick Vamivakas},
title = {Optomechanics with levitated particles},
journal = {Rep. Prog. Phys.}
}

@article{Cernotik2020,
  title = {Strong mechanical squeezing for a levitated particle by coherent scattering},
  author={{\v{C}}ernot{\'\i}k, Ond{\v{r}}ej and Filip, Radim},
  journal = {Phys. Rev. Res.},
  volume = {2},
  issue = {1},
  pages = {013052},
  numpages = {8},
  year = {2020},
  month = {1},
  publisher = {American Physical Society},
  doi = {10.1103/PhysRevResearch.2.013052},
  url = {https://link.aps.org/doi/10.1103/PhysRevResearch.2.013052}
}

@article{Watson_2020,
title = {Normal modes for N identical particles: A study of the evolution of collective behavior from few-body to many-body},
journal = {Annals of Physics},
volume = {419},
pages = {168219},
year = {2020},
issn = {0003-4916},
doi = {https://doi.org/10.1016/j.aop.2020.168219},
url = {https://www.sciencedirect.com/science/article/pii/S0003491620301536},
author = {D.K. Watson}
}

@article{Mason2019,
author={Mason, David
and Chen, Junxin
and Rossi, Massimiliano
and Tsaturyan, Yeghishe
and Schliesser, Albert},
title={Continuous force and displacement measurement below the standard quantum limit},
journal={Nat. Phys.},
year={2019},
month={8},
day={01},
volume={15},
number={8},
pages={745-749},
issn={1745-2481},
doi={10.1038/s41567-019-0533-5},
url={https://doi.org/10.1038/s41567-019-0533-5}
}

@article{Winstone2022,
  title = {Optical Trapping of High-Aspect-Ratio NaYF Hexagonal Prisms for kHz-MHz Gravitational Wave Detectors},
  author = {Winstone, George and Wang, Zhiyuan and Klomp, Shelby and Felsted, Robert G. and Laeuger, Andrew and Gupta, Chaman and Grass, Daniel and Aggarwal, Nancy and Sprague, Jacob and Pauzauskie, Peter J. and Larson, Shane L. and Kalogera, Vicky and Geraci, Andrew A.},
  collaboration = {LSD Collaboration},
  journal = {Phys. Rev. Lett.},
  volume = {129},
  issue = {5},
  pages = {053604},
  numpages = {7},
  year = {2022},
  month = {7},
  publisher = {American Physical Society},
  doi = {10.1103/PhysRevLett.129.053604},
  url = {https://link.aps.org/doi/10.1103/PhysRevLett.129.053604}
}

@article{G-Ballestero2021rev,
author = {C. Gonzalez-Ballestero  and M. Aspelmeyer  and L. Novotny  and R. Quidant  and O. Romero-Isart },
title = {Levitodynamics: Levitation and control of microscopic objects in vacuum},
journal = {Science},
volume = {374},
number = {6564},
pages = {eabg3027},
year = {2021},
doi = {10.1126/science.abg3027},
url = {https://www.science.org/doi/abs/10.1126/science.abg3027},
eprint = {https://www.science.org/doi/pdf/10.1126/science.abg3027}
}

@article{Piotrowski2023,
author={Piotrowski, Johannes
and Windey, Dominik
and Vijayan, Jayadev
and Gonzalez-Ballestero, Carlos
and de los R{\'i}os Sommer, Andr{\'e}s
and Meyer, Nadine
and Quidant, Romain
and Romero-Isart, Oriol
and Reimann, Ren{\'e}
and Novotny, Lukas},
title={Simultaneous ground-state cooling of two mechanical modes of a levitated nanoparticle},
journal={Nat. Phys.},
year={2023},
month={7},
day={01},
volume={19},
number={7},
pages={1009-1013},
issn={1745-2481},
doi={10.1038/s41567-023-01956-1},
url={https://doi.org/10.1038/s41567-023-01956-1}
}

@article{Moore2021rev,
doi = {10.1088/2058-9565/abcf8a},
url = {https://dx.doi.org/10.1088/2058-9565/abcf8a},
year = {2021},
month = {1},
publisher = {IOP Publishing},
volume = {6},
number = {1},
pages = {014008},
author = {David C Moore and Andrew A Geraci},
title = {Searching for new physics using optically levitated sensors},
journal = {Quantum Sci. Technol.}
}

@article{penny2021performance,
  title={Performance and limits of feedback cooling methods for levitated oscillators: A direct comparison},
  author={Penny, TW and Pontin and A and Barker, PF},
  journal={Phys. Rev. A},
  volume={104},
  number={2},
  pages={023502},
  year={2021},
  publisher={APS}
}

@article{tebbenjohanns2021quantum,
  title={Quantum control of a nanoparticle optically levitated in cryogenic free space},
  author={Tebbenjohanns, Felix and Mattana, M Luisa and Rossi, Massimiliano and Frimmer, Martin and Novotny, Lukas},
  journal={Nature},
  volume={595},
  number={7867},
  pages={378--382},
  year={2021},
  publisher={Nature Publishing Group UK London}
}

@article{ren2022event,
  title={Event-based imaging of levitated microparticles},
  author={Ren, Yugang and Benedetto, Enrique and Borrill, Harry and Savchuk, Yelizaveta and O'Flynn, Katie and Rashid, Muddassar and Millen, James and others},
  journal={Appl. Phys. Lett.},
  volume={121},
  number={11},
  year={2022},
  publisher={AIP Publishing}
}

@article{piotrowski2023simultaneous,
  title={Simultaneous ground-state cooling of two mechanical modes of a levitated nanoparticle},
  author={Piotrowski, Johannes and Windey, Dominik and Vijayan, Jayadev and Gonzalez-Ballestero, Carlos and de los R{\'\i}os Sommer, Andr{\'e}s and Meyer, Nadine and Quidant, Romain and Romero-Isart, Oriol and Reimann, Ren{\'e} and Novotny, Lukas},
  journal={Nat. Phys.},
  volume={19},
  number={7},
  pages={1009--1013},
  year={2023},
  publisher={Nature Publishing Group UK London}
}

@article{ranjit2016zeptonewton,
  title={Zeptonewton force sensing with nanospheres in an optical lattice},
  author={Ranjit, Gambhir and Cunningham, Mark and Casey, Kirsten and Geraci, Andrew A},
  journal={Phys. Rev. A},
  volume={93},
  number={5},
  pages={053801},
  year={2016},
  publisher={APS}
}

@article{monteiro2020force,
  title={Force and acceleration sensing with optically levitated nanogram masses at microkelvin temperatures},
  author={Monteiro, Fernando and Li, Wenqiang and Afek, Gadi and Li, Chang-ling and Mossman, Michael and Moore, David C},
  journal={Phys. Rev. A},
  volume={101},
  number={5},
  pages={053835},
  year={2020},
  publisher={APS}
}

@article{bose2017spin,
  title={Spin entanglement witness for quantum gravity},
  author={Bose, Sougato and Mazumdar, Anupam and Morley, Gavin W and Ulbricht, Hendrik and Toro{\v{s}}, Marko and Paternostro, Mauro and Geraci, Andrew A and Barker, Peter F and Kim, MS and Milburn, Gerard},
  journal={Phys. Rev. Lett.},
  volume={119},
  number={24},
  pages={240401},
  year={2017},
  publisher={APS}
}

@article{marletto2017gravitationally,
  title={Gravitationally induced entanglement between two massive particles is sufficient evidence of quantum effects in gravity},
  author={Marletto, Chiara and Vedral, Vlatko},
  journal={Phys. Rev. Lett.},
  volume={119},
  number={24},
  pages={240402},
  year={2017},
  publisher={APS}
}

@article{arvanitaki2013detecting,
  title={Detecting high-frequency gravitational waves with optically levitated sensors},
  author={Arvanitaki, Asimina and Geraci, Andrew A},
  journal={Phys. Rev. Lett.},
  volume={110},
  number={7},
  pages={071105},
  year={2013},
  publisher={APS}
}

@article{rieser2022tunable,
  title={Tunable light-induced dipole-dipole interaction between optically levitated nanoparticles},
  author={Rieser, Jakob and Ciampini, Mario A and Rudolph, Henning and Kiesel, Nikolai and Hornberger, Klaus and Stickler, Benjamin A and Aspelmeyer, Markus and Deli{\'c}, Uro{\v{s}}},
  journal={Science},
  volume={377},
  number={6609},
  pages={987--990},
  year={2022},
  publisher={American Association for the Advancement of Science}
}

@article{livska2023cold,
  title={Cold damping of levitated optically coupled nanoparticles},
  author={Li{\v{s}}ka, Vojt{\v{e}}ch and Zem{\'a}nkov{\'a}, Tereza and Svak, Vojt{\v{e}}ch and J{\'a}kl, Petr and Je{\v{z}}ek, Jan and Br{\'a}neck{\`y}, Martin and Simpson, Stephen H and Zem{\'a}nek, Pavel and Brzobohat{\`y}, Oto},
  journal={Optica},
  volume={10},
  number={9},
  pages={1203--1209},
  year={2023},
  publisher={Optica Publishing Group}
}

@article{Liao2021,
doi = {10.1088/1674-4926/42/1/013105},
url = {https://dx.doi.org/10.1088/1674-4926/42/1/013105},
year = {2021},
month = {1},
publisher = {Chinese Institute of Electronics},
volume = {42},
number = {1},
pages = {013105},
author = {Fuyou Liao and Feichi  and Zhou and Yang Chai},
title = {Neuromorphic vision sensors: Principle, progress and perspectives},
journal = {J. Semicond.}
}

@article{Vanarse2016,
AUTHOR={Vanarse, Anup  and Osseiran and Adam  and Rassau, Alexander },
TITLE={A Review of Current Neuromorphic Approaches for Vision, Auditory, and Olfactory Sensors},
JOURNAL={Front. Neurosci.},
VOLUME={10},
YEAR={2016},
url={https://www.frontiersin.org/journals/neuroscience/articles/10.3389/fnins.2016.00115},
doi={10.3389/fnins.2016.00115},
ISSN={1662-453X}
}

@inproceedings{Ng2022,
  title={Asynchronous kalman filter for event-based star tracking},
  author={Ng, Yonhon and Latif, Yasir and Chin, Tat-Jun and Mahony, Robert},
  booktitle={European Conference on Computer Vision},
  pages={66-79},
  year={2022},
  organization={Springer}
}

@article{Werneck2024,
    author = {Werneck, L. R. and Jessup, C. and Brandenberger, A. and Knowles, T. and Lewandowski, C. W. and Nolan, M. and Sible, K. and Etienne, Z. B. and D’Urso, B.},
    title = {Cross-correlation image analysis for real-time single particle tracking},
    journal = {Rev. Sci. Instrum.},
    volume = {95},
    number = {7},
    pages = {073708},
    year = {2024},
    month = {07},
	issn = {0034-6748},
    doi = {10.1063/5.0206405},
    url = {https://doi.org/10.1063/5.0206405},
    eprint = {https://pubs.aip.org/aip/rsi/article-pdf/doi/10.1063/5.0206405/20050243/073708\_1\_5.0206405.pdf}
}

@inproceedings{Sikorski2021,
	author = {Sikorski, O. and Izzo and D. and Gabriele Meoni},
	year = {2021},
	month = {6},
	pages = {1941--1950},
	booktitle = {Proceedings of the IEEE/CVF Conference on Computer Vision},
        title={Event-based spacecraft landing using time-to-contact}
}

@article{arita2022all,
  title={All-optical sub-Kelvin sympathetic cooling of a levitated microsphere in vacuum},
  author={Arita, Yoshihiko and Bruce, Graham David and Wright, Ewan Malcolm and Simpson, Stephen H and Zem{\'a}nek, Pavel and Dholakia, Kishan},
  journal={Optica},
  volume={9},
  number={9},
  pages={1000--1002},
  year={2022},
  publisher={Optica Publishing Group}
}

@article{arita2018optical,
  title={Optical binding of two cooled micro-gyroscopes levitated in vacuum},
  author={Arita, Yoshihiko and Wright and Ewan M and Dholakia and Kishan},
  journal={Optica},
  volume={5},
  number={8},
  pages={910--917},
  year={2018},
  publisher={Optica Publishing Group}
}

@article{bykov2019direct,
  title={Direct loading of nanoparticles under high vacuum into a Paul trap for levitodynamical experiments},
  author={Bykov, Dmitry S and Mestres, Pau and Dania, Lorenzo and Schm{\"o}ger, Lisa and Northup, Tracy E},
  journal={Appl. Phys. Lett.},
  volume={115},
  pages = {034101},
  number={3},
  year={2019},
  publisher={AIP Publishing}
}

@article{Bullier2019,
    author = {Bullier, N. P. and Pontin and A. and Barker, P. F.},
    title = {Super-resolution imaging of a low frequency levitated oscillator},
    journal = {Rev. Sci. Instrum.},
    volume = {90},
    number = {9},
    pages = {093201},
    year = {2019},
    month = {09},
    issn = {0034-6748},
    doi = {10.1063/1.5108807},
    url = {https://doi.org/10.1063/1.5108807},
    eprint = {https://pubs.aip.org/aip/rsi/article-pdf/doi/10.1063/1.5108807/15612939/093201\_1\_online.pdf}
}

@article{Palit1997,
title = {Signal extraction from multiple noisy sensors},
journal = {Signal Processing},
volume = {61},
number = {3},
pages = {199-212},
year = {1997},
issn = {0165-1684},
doi = {https://doi.org/10.1016/S0165-1684(97)00105-9},
url = {https://www.sciencedirect.com/science/article/pii/S0165168497001059},
author = {Sarbani Palit}
}

@article{Sasiadek2002,
title = {Sensor fusion},
journal = {Annu. Rev. Control},
volume = {26},
number = {2},
pages = {203-228},
year = {2002},
issn = {1367-5788},
doi = {https://doi.org/10.1016/S1367-5788(02)00045-7},
url = {https://www.sciencedirect.com/science/article/pii/S1367578802000457},
author = {J.Z. Sasiadek}
}

@article{Nazarahari2021,
title = {40 years of sensor fusion for orientation tracking via magnetic and inertial measurement units: Methods, lessons learned, and future challenges},
journal = {Inf. Fusion},
volume = {68},
pages = {67-84},
year = {2021},
issn = {1566-2535},
doi = {https://doi.org/10.1016/j.inffus.2020.10.018},
url = {https://www.sciencedirect.com/science/article/pii/S1566253520303997},
author = {Milad Nazarahari and Hossein Rouhani}
}

@article{chang2010,
	author = {Chang, D.E. and Regal, C.A. and Papp, S.B. and Wilson, D.J. and Ye, J. and Painter, O. and Kimble, H.J. and Zoller, P.},
	journal = {Proceedings of the National Academy of Sciences},
	number = {3},
	pages = {1005--1010},
	publisher = {National Acad Sciences},
	title = {Cavity opto-mechanics using an optically levitated nanosphere},
	volume = {107},
	year = {2010}
}

@article{barker2010,
  title={Cavity cooling of an optically trapped nanoparticle},
  author={Barker, PF and Shneider, MN},
  journal={Phys. Rev. A},
  volume={81},
  number={2},
  pages={023826},
  year={2010},
  publisher={APS}
}

@article{romeroisart2010,
  title={Toward quantum superposition of living organisms},
  author={Romero-Isart, Oriol and Juan, Mathieu L and Quidant, Romain and Cirac, J Ignacio},
  journal={New J. Phys.},
  volume={12},
  number={3},
  pages={033015},
  year={2010},
  publisher={IOP Publishing}
}

@article{Jain2016,
  title = {Direct Measurement of Photon Recoil from a Levitated Nanoparticle},
  author = {Jain, Vijay and Gieseler, Jan and Moritz, Clemens and Dellago, Christoph and Quidant, Romain and Novotny, Lukas},
  journal = {Phys. Rev. Lett.},
  volume = {116},
  issue = {24},
  pages = {243601},
  numpages = {5},
  year = {2016},
  month = {6},
  publisher = {American Physical Society},
  doi = {10.1103/PhysRevLett.116.243601},
  url = {http://link.aps.org/doi/10.1103/PhysRevLett.116.243601}
}

@article{Vovrosh2017,
author = {Jamie Vovrosh and Muddassar Rashid and David Hempston and James Bateman and Mauro Paternostro and Hendrik Ulbricht},
journal = {J. Opt. Soc. Am. B},
number = {7},
pages = {1421--1428},
publisher = {Optica Publishing Group},
title = {Parametric feedback cooling of levitated optomechanics in a parabolic mirror trap},
volume = {34},
month = {7},
year = {2017},
url = {https://opg.optica.org/josab/abstract.cfm?URI=josab-34-7-1421},
doi = {10.1364/JOSAB.34.001421}
}

@article{Hupfl2023,
  title = {Optimal Cooling of Multiple Levitated Particles through Far-Field Wavefront Shaping},
  author = {H\"upfl, Jakob and Bachelard, Nicolas and Kaczvinszki, Markus and Horodynski, Michael and K\"uhmayer, Matthias and Rotter, Stefan},
  journal = {Phys. Rev. Lett.},
  volume = {130},
  issue = {8},
  pages = {083203},
  numpages = {7},
  year = {2023},
  month = {2},
  publisher = {American Physical Society},
  doi = {10.1103/PhysRevLett.130.083203},
  url = {https://link.aps.org/doi/10.1103/PhysRevLett.130.083203}
}

@article{Tebbenjohanns2019,
  title = {Cold Damping of an Optically Levitated Nanoparticle to Microkelvin Temperatures},
  author = {Tebbenjohanns, Felix and Frimmer, Martin and Militaru, Andrei and Jain, Vijay and Novotny, Lukas},
  journal = {Phys. Rev. Lett.},
  volume = {122},
  issue = {22},
  pages = {223601},
  numpages = {6},
  year = {2019},
  month = {6},
  publisher = {American Physical Society},
  doi = {10.1103/PhysRevLett.122.223601},
  url = {https://link.aps.org/doi/10.1103/PhysRevLett.122.223601}
}

@article{Conangla2019,
  title = {Optimal Feedback Cooling of a Charged Levitated Nanoparticle with Adaptive Control},
  author = {Conangla, Gerard P. and Ricci, Francesco and Cuairan, Marc T. and Schell, Andreas W. and Meyer, Nadine and Quidant, Romain},
  journal = {Phys. Rev. Lett.},
  volume = {122},
  issue = {22},
  pages = {223602},
  numpages = {6},
  year = {2019},
  month = {6},
  publisher = {American Physical Society},
  doi = {10.1103/PhysRevLett.122.223602},
  url = {https://link.aps.org/doi/10.1103/PhysRevLett.122.223602}
}

@article{Kane2010,
  title = {Levitated spinning graphene flakes in an electric quadrupole ion trap},
  author = {Kane, B. E.},
  journal = {Phys. Rev. B},
  volume = {82},
  issue = {11},
  pages = {115441},
  numpages = {13},
  year = {2010},
  month = {9},
  publisher = {American Physical Society},
  doi = {10.1103/PhysRevB.82.115441},
  url = {https://link.aps.org/doi/10.1103/PhysRevB.82.115441}
}

@article{Kuhlicke2014,
    author = {Kuhlicke, Alexander and Schell, Andreas W. and Zoll, Joachim and Benson, Oliver},
    title = {Nitrogen vacancy center fluorescence from a submicron diamond cluster levitated in a linear quadrupole ion trap},
    journal = {Appl. Phys. Lett.},
    volume = {105},
    number = {7},
    pages = {073101},
    year = {2014},
    month = {08},
	issn = {0003-6951},
    doi = {10.1063/1.4893575},
    url = {https://doi.org/10.1063/1.4893575},
    eprint = {https://pubs.aip.org/aip/apl/article-pdf/doi/10.1063/1.4893575/14316710/073101\_1\_online.pdf}
}

@article{Liang2022,
  title    = "Yoctonewton force detection based on optically levitated
              oscillator",
  author   = "Liang, Tao and Zhu, Shaochong and He, Peitong and Chen, Zhiming
              and Wang, Yingying and Li, Cuihong and Fu, Zhenhai and Gao,
              Xiaowen and Chen, Xinfan and Li, Nan and Zhu, Qi and Hu, Huizhu",
  journal  = "Fundam Res",
  volume   =  3,
  number   =  1,
  pages    = "57--62",
  month    =  10,
  year     =  2022,
  address  = "Netherlands",
  language = "en"
}

@article{Millen2015,
  title = {Cavity Cooling a Single Charged Levitated Nanosphere},
  author = {Millen, J. and Fonseca, P. Z. G. and Mavrogordatos, T. and Monteiro, T. S. and Barker, P. F.},
  journal = {Phys. Rev. Lett.},
  volume = {114},
  issue = {12},
  pages = {123602},
  numpages = {5},
  year = {2015},
  month = {3},
  publisher = {American Physical Society},
  doi = {10.1103/PhysRevLett.114.123602},
  url = {https://link.aps.org/doi/10.1103/PhysRevLett.114.123602}
}

@article{Delord2017,
doi = {10.1088/1367-2630/aa659c},
url = {https://dx.doi.org/10.1088/1367-2630/aa659c},
year = {2017},
month = {3},
publisher = {IOP Publishing},
volume = {19},
number = {3},
pages = {033031},
author={Delord, Tom and Nicolas, Louis and Schwab, Lucien and H{\'e}tet, Gabriel},
title = {Electron spin resonance from NV centers in diamonds levitating in an ion trap},
journal = {New J. Phys.}
}

@article{Poggio2007,
  title = {Feedback Cooling of a Cantilever's Fundamental Mode below 5 mK},
  author = {Poggio, M. and Degen, C. L. and Mamin, H. J. and Rugar, D.},
  journal = {Phys. Rev. Lett.},
  volume = {99},
  issue = {1},
  pages = {017201},
  numpages = {4},
  year = {2007},
  month = {7},
  publisher = {American Physical Society},
  doi = {10.1103/PhysRevLett.99.017201},
  url = {https://link.aps.org/doi/10.1103/PhysRevLett.99.017201}
}

@article{pontin2020ultranarrow,
  title={Ultranarrow-linewidth levitated nano-oscillator for testing dissipative wave-function collapse},
  author={Pontin, A. and Bullier, NP. and Toro{\v{s}}, M. and Barker, PF.},
  journal={Phys. Rev. Res.},
  volume={2},
  number={2},
  pages={023349},
  year={2020},
  publisher={APS}
}

@article{magrini2021real,
  title={Real-time optimal quantum control of mechanical motion at room temperature},
  author={Magrini, Lorenzo and Rosenzweig, Philipp and Bach, Constanze and Deutschmann-Olek, Andreas and Hofer, Sebastian G and Hong, Sungkun and Kiesel, Nikolai and Kugi, Andreas and Aspelmeyer, Markus},
  journal={Nature},
  volume={595},
  number={7867},
  pages={373--377},
  year={2021},
  publisher={Nature Publishing Group UK London}
}

@article{kamba2022optical,
  title={Optical cold damping of neutral nanoparticles near the ground state in an optical lattice},
  author={Kamba, Mitsuyoshi and Shimizu and Ryoga and Aikawa, Kiyotaka},
  journal={Optics Express},
  volume={30},
  number={15},
  pages={26716--26727},
  year={2022},
  publisher={Optica Publishing Group}
}

@article{ranfagni2022two,
  title={Two-dimensional quantum motion of a levitated nanosphere},
  author={Ranfagni, A and B{\o}rkje, Kjetil and Marino, F and Marin, F},
  journal={Physical Review Research},
  volume={4},
  number={3},
  pages={033051},
  year={2022},
  publisher={APS}
}

@article{Sharma2024,
author={Sharma, Deepak
and Rath, Santi Prasad
and Kundu, Bidyabhusan
and Korkmaz, Anil
and S, Harivignesh
and Thompson, Damien
and Bhat, Navakanta
and Goswami, Sreebrata
and Williams, R. Stanley
and Goswami, Sreetosh},
title={Linear symmetric self-selecting 14-bit kinetic molecular memristors},
journal={Nature},
year={2024},
month={9},
day={01},
volume={633},
number={8030},
pages={560-566},
issn={1476-4687},
doi={10.1038/s41586-024-07902-2},
url={https://doi.org/10.1038/s41586-024-07902-2}
}

@article{Ambrogio2018,
author={Ambrogio, Stefano
and Narayanan, Pritish
and Tsai, Hsinyu
and Shelby, Robert M.
and Boybat, Irem
and di Nolfo, Carmelo
and Sidler, Severin
and Giordano, Massimo
and Bodini, Martina
and Farinha, Nathan C. P.
and Killeen, Benjamin
and Cheng, Christina
and Jaoudi, Yassine
and Burr, Geoffrey W.},
title={Equivalent-accuracy accelerated neural-network training using analogue memory},
journal={Nature},
year={2018},
month={6},
day={01},
volume={558},
number={7708},
pages={60-67},
issn={1476-4687},
doi={10.1038/s41586-018-0180-5},
url={https://doi.org/10.1038/s41586-018-0180-5}
}

@article{Xiao2020,
    author = {Xiao, T. Patrick and Bennett, Christopher H. and Feinberg, Ben and Agarwal, Sapan and Marinella, Matthew J.},
    title = {Analog architectures for neural network acceleration based on non-volatile memory},
    journal = {Applied Physics Reviews},
    volume = {7},
    number = {3},
    pages = {031301},
    year = {2020},
    month = {07},
	issn = {1931-9401},
    doi = {10.1063/1.5143815},
    url = {https://doi.org/10.1063/1.5143815},
    eprint = {https://pubs.aip.org/aip/apr/article-pdf/doi/10.1063/1.5143815/19740151/031301\_1\_online.pdf}
}

@article{Furmonas2022,
AUTHOR = {Furmonas, Justas and Liobe, John and Barzdenas and Vaidotas},
TITLE = {Analytical Review of Event-Based Camera Depth Estimation Methods and Systems},
JOURNAL = {Sensors},
VOLUME = {22},
YEAR = {2022},
NUMBER = {3},
ARTICLE-NUMBER = {1201},
URL = {https://www.mdpi.com/1424-8220/22/3/1201},
PubMedID = {35161946},
ISSN = {1424-8220},
DOI = {10.3390/s22031201}
}

@book{penny2000signal,
  title={Signal processing course},
  author={Penny, William D},
  journal={Institute of Neurology, University College London, UK},
  year={2000},
  publisher={Citeseer}
}

@ARTICLE{Apps2025,
       author = {{Apps}, Angus and {Wang}, Ziwei and {Perejogin}, Vladimir and {Molloy}, Timothy and {Mahony}, Robert},
        title = "{Asynchronous Multi-Object Tracking with an Event Camera}",
      journal = {arXiv e-prints},
         year = 2025,
        month = may,
          eid = {arXiv:2505.08126},
        pages = {arXiv:2505.08126},
          doi = {10.48550/arXiv.2505.08126},
archivePrefix = {arXiv},
       eprint = {2505.08126},
 primaryClass = {cs.CV},
       adsurl = {https://ui.adsabs.harvard.edu/abs/2025arXiv250508126A}
}

@InProceedings{Kim2016,
author="Kim, Hanme
and Leutenegger and Stefan
and Davison, Andrew J.",
editor="Leibe, Bastian
and Matas, Jiri
and Sebe, Nicu
and Welling, Max",
title="Real-Time 3D Reconstruction and 6-DoF Tracking with an Event Camera",
booktitle="Computer Vision -- ECCV 2016",
year="2016",
publisher="Springer International Publishing",
address="Cham",
pages="349--364",
isbn="978-3-319-46466-4"
}

@article{obata2010multi,
  title={Multi-focus two-photon polymerization technique based on individually controlled phase modulation},
  author={Obata, Kotaro and Koch, J{\"u}rgen and Hinze, Ulf and Chichkov, Boris N},
  journal={Optics express},
  volume={18},
  number={16},
  pages={17193--17200},
  year={2010},
  publisher={Optical Society of America}
}
\end{refsection}

\end{document}